\documentclass[10pt,twocolumn,twoside]{IEEEtran}

\usepackage{romannum}
\usepackage{amssymb}
\usepackage{amsmath,amsthm}
\usepackage{graphicx}
\usepackage{dblfloatfix} 
\usepackage{multicol}
\usepackage{lipsum}
\usepackage[linesnumbered, ruled, vlined]{algorithm2e}
\usepackage{booktabs}
\usepackage{tabularx}
\usepackage{ragged2e}
\usepackage{csquotes}
\usepackage{helvet}
\usepackage{textcomp}
\usepackage[T1]{fontenc}
\usepackage{bm}
\usepackage{verbatim} 
\usepackage{mathtools}
\usepackage{caption}
\usepackage{graphicx} 
\usepackage{threeparttable}
\usepackage{textcomp}
\usepackage{commath}
\usepackage{float}
\usepackage{svg}
\usepackage{tabularx}
\usepackage{amsmath}
\usepackage{color}
\usepackage{bm}
\usepackage{bbm}
\usepackage{mathrsfs,float} 
\usepackage{epstopdf}
\usepackage{float}
\usepackage{gensymb,stmaryrd}
\usepackage{algorithmicx}
 \usepackage{enumerate}
 \usepackage{enumitem}
  \newcolumntype{P}[1]{>{\centering\arraybackslash}p{#1}}
  \newcolumntype{M}[1]{>{\centering\arraybackslash}m{#1}}

\usepackage{hyperref}
\usepackage{textcomp}
\usepackage[normalem]{ulem}
\usepackage{xr}

\usepackage{romannum}
\usepackage{amssymb}
\usepackage{rotating} 
\usepackage{breqn}

\usepackage{color}
\definecolor{lightgray}{gray}{0.95}

\usepackage{hyperref}

\usepackage{xr}
\makeatletter

\newcommand*{\addFileDependency}[1]{
\typeout{(#1)}
\@addtofilelist{#1}
\IfFileExists{#1}{}{\typeout{No file #1.}}
}\makeatother

\usepackage{booktabs, multirow} 
\usepackage{soul}

\usepackage{caption}
\usepackage{subcaption}
\usepackage{cite}

\ifCLASSINFOpdf
\else
\fi
\usepackage{array}
\usepackage{fixltx2e}
\usepackage{tabularray}
\usepackage{url}
\usepackage{pifont}

\DeclareUnicodeCharacter{2061}{}

\begin{document}
%
\title{Neural Signal Compression using RAMAN tinyML Accelerator for BCI Applications}

%

\author{Adithya Krishna, Sohan Debnath, Madhuvanthi Srivatsav, Andr\'e van Schaik, Mahesh Mehendale and Chetan Singh Thakur*
\thanks{A. Krishna is with the Department of Electronic Systems Engineering, Indian Institute of Science, Bangalore - 560012, India, and also with the International Centre for Neuromorphic Systems, The MARCS Institute, Western Sydney University, Australia. \\ 
S. Debnath, M. Srivatsav, M. Mehendale, and C. S. Thakur (Email: csthakur@iisc.ac.in) are with the Department of Electronic Systems Engineering, Indian Institute of Science, Bangalore - 560012, India. \\ 
A. van Schaik is with the International Centre for Neuromorphic Systems, The MARCS Institute, Western Sydney University, Australia.
\newline This work was supported by the Pratiksha Trust grant BCD - FG/SMCH-22-2106 and INAE grant INAE/121/AKF/48 (SAP code - SP/INAE-23-0001).
\newline *Corresponding author}} 
\maketitle

\begin{abstract}
High-quality, multi-channel neural recording is indispensable for neuroscience research and clinical applications. Large-scale brain recordings often produce vast amounts of data that must be wirelessly transmitted for subsequent offline analysis and decoding, especially in brain-computer interfaces (BCIs) utilizing high-density intracortical recordings with hundreds or thousands of electrodes. However, transmitting raw neural data presents significant challenges due to limited communication bandwidth and resultant excessive heating. To address this challenge, we propose a neural signal compression scheme utilizing Convolutional Autoencoders (CAEs), which achieves a compression ratio of up to 150 for compressing local field potentials (LFPs). The CAE encoder section is implemented on RAMAN, an energy-efficient tinyML accelerator designed for edge computing. RAMAN leverages sparsity in activation and weights through zero skipping, gating, and weight compression techniques. Additionally, we employ hardware-software co-optimization by pruning the CAE encoder model parameters using a hardware-aware balanced stochastic pruning strategy, resolving workload imbalance issues and eliminating indexing overhead to reduce parameter storage requirements by up to 32.4\%. Post layout simulation shows that the RAMAN encoder can be implemented in a TSMC 65-nm CMOS process, occupying a core area of 0.0187~mm\textsuperscript{2} per channel. Operating at a clock frequency of 2~MHz and a supply voltage of 1.2 V, the estimated power consumption is 15.1~$\mu$W per channel for the proposed DS-CAE1 model. For functional validation, the RAMAN encoder was also deployed on an Efinix Ti60 FPGA, utilizing 37.3k LUTs and 8.6k flip-flops. The compressed neural data from RAMAN is reconstructed offline with signal-to-noise and distortion ratios (SNDR) of 22.6 dB and 27.4 dB, along with R2 scores of 0.81 and 0.94, respectively, evaluated on two monkey neural recordings.

\end{abstract}


\section{Introduction}
In recent years, the Brain-Computer Interface (BCI) has garnered significant attention for facilitating direct communication between the human brain and external devices \cite{Wolpaw2002,NicolasAlonso2012,Chaudhary2016,Demarest2024}. BCIs have emerged as a revolutionary tool for advancing our understanding of the brain and are increasingly being utilized across various clinical applications \cite{Josephclinical2009,shih2012brain,Krishna2024biocas}, providing inventive solutions for communication \cite{McFarland2011}, control \cite{Karas2023,Wolpaw2002}, and rehabilitation \cite{Cervera2018,Romagosa2020,Mane2020,Baniqued2021}. Ongoing improvements in signal processing, machine learning algorithms, and neurotechnology pave the way for BCIs to revolutionize healthcare, human-computer interaction, and beyond. BCIs can potentially restore lost sensory abilities, such as vision and hearing \cite{Hwang2015,Niketeghad2019,Wang2023} through stimulation and regain lost motor functions for individuals with motor impairments, such as those caused by conditions like ALS or spinal cord injury \cite{Vansteensel2016,Chaudhary2017}. In clinical settings, BCIs can interpret the user's intentions from brain activity and use this information to control the person's limb or an assistive device, such as a prosthetic arm or a computer cursor, necessitating simultaneous recordings from a relatively large population of neurons to achieve acceptable prediction accuracy. Intracortical neural recording systems have evolved significantly, progressing from widely used recording arrays like the Utah Array \cite{Harrison2008}, which supports up to a hundred recording sites, to high-density electrodes such as \cite{Oxley2016minimally}, Neuralink \cite{Elon2019}, and Neuropixel \cite{Steinmetz2021}. These state-of-the-art intracortical neural recording systems have thousands of recording sites, producing massive amounts of neural data that require wireless transmission for offline signal analysis and decoding.

\par 
Continuous wireless transmission of raw neural data poses a significant challenge due to constrained communication bandwidth (low data rates) and leads to excessive heating. For instance, a system comprising 1024 channels, sampled at 30 kS/s with 16 bits per sample, generates 491.52 Mb/s of data that must be wirelessly transmitted. To mitigate this, the data undergoes compression before transmission. Compressed sensing (CS) is one such prominent method that reduces the data dimensionality by multiplying the input signal $X$ of dimension $(1 \times M)$ with a sensing matrix of size $(M \times N)$ to produce the compressed output signal $Y$ of dimension $(1 \times N)$, with compression ratio (CR) $M/N (M>>N)$ \cite{Wang2015, Shrivastwa2020}. Reconstruction of the compressed signal to its original form utilizes CS-based reconstruction algorithms \cite{Shoaran2014}, with an emerging interest in deep learning-based methods \cite{Shrivastwa2020}.
Additionally, various lossless compression techniques, categorized into dictionary-based and statistical-based schemes, offer further compression options for neural data. Dictionary-based compression \cite{Delgado2003,Daou2014,Vadori2016,Qian2020} relies on creating a dictionary of frequently occurring patterns in the data and replacing them with shorter codes or references to entries in the dictionary. Statistical compression utilizes statistical models to represent and encode data more efficiently.  One standard method of statistical compression is Huffman coding \cite{Bihr2014,Savolainen2020}. In Huffman coding, symbols with higher frequencies are assigned shorter codewords, while symbols with lower frequencies are assigned longer codewords. This ensures that more common symbols are represented with fewer bits, leading to overall compression. Lossless compression schemes like Huffman coding ensure that the original input signal is accurately reconstructed from the compressed data without any reconstruction errors. In contrast, lossy compression methods, such as compressed sensing (CS), can achieve significantly higher compression ratios but at the expense of introducing reconstruction errors. Additionally, several spike compression schemes have been proposed \cite{Shaeri2020,Chen2023,Shaeri2015,Mohan2023}, with associated data transformations to better represent the signal space \cite{Shaeri2023}. These methods focus exclusively on compressing spike waveforms for applications employing single-unit activities (SUAs) and multi-unit activities (MUAs). In contrast, our work focuses on compressing local field potentials (LFPs), where spike detection, extraction, and sorting methods are not applicable. LFPs, which capture slower voltage fluctuations (<300 Hz) with lower spatial resolution than SUAs and MUAs, offer greater long-term stability against factors such as electrode drift, neuron drop-out, and scarring at the electrode-tissue interface \cite{Salatino2018}. While MUA-based BCIs primarily decode motor cortex activity, LFP-based BCIs target higher-level cognitive regions like the posterior parietal cortex, enabling advanced cognitive decoding \cite{Musallam2004}. Studies have shown that LFP signals can reliably decode neural activity; for example, \cite{Aflalo2015} demonstrated that movement intentions, kinematic trajectories, and movement types can be accurately predicted from LFP signals.
\par 
In this work, we propose a convolutional autoencoder-based neural signal compression scheme for compressing the LFPs. Autoencoders are neural networks used for the unsupervised learning of efficient codings or representations of the input data \cite{Baldi2011,Zhai2018,Li2023,Bank2023}. They consist of two main components: an encoder and a decoder. The encoder network maps the input data to a lower-dimensional latent space representation, also known as a codeword or encoding, which effectively compresses the neural signal. The decoder network takes the encoded representation produced by the encoder and attempts to reconstruct the original input signal from it. The encoder network is implemented using RAMAN, a {\emph{R}}e-configurable and sp{\emph{A}}rse {tiny{\emph{M}}L} {\emph{A}}ccelerator for infere{\emph{N}}ce proposed in \cite{RAMAN2023} for edge computing applications to compress the input data to a latent space.  
\begin{figure*}[t]
  \centering
    \includegraphics[width=18cm]{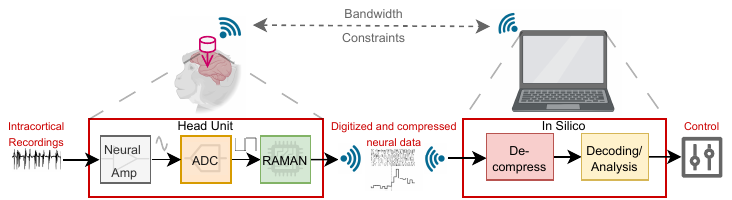}
  \caption{A system-level overview of the proposed neural signal compression scheme utilizing convolutional autoencoders.}
  \label{fig:top_level_illust}
\end{figure*}

\subsection{Our Contributions}
The main contributions of this paper can be summarized as follows:
\begin{enumerate}
    \item We propose a neural signal compression scheme using convolutional autoencoders (CAEs) that achieves superior compression ratios compared to existing LFP compression methods, as detailed in Section \ref{sec:proposed_scheme} and Section \ref{sec:mod_performance}. We evaluate the performance of various CAE architectures based on SNDR and R2 scores, employing depth-wise separable convolutions to reduce Multiply-Accumulate (MAC) operations.

    \item We employ RAMAN, a low-power, compact tinyML accelerator designed for edge computing \cite{RAMAN2023}, to implement the CAE encoder. RAMAN leverages sparsity in activations and weights to reduce latency, memory storage, and power consumption. We assess RAMAN's performance with MobileNetV1-based convolutional autoencoder models and introduce our custom, compact models, DS-CAE1 and DS-CAE2 (Depthwise Separable Convolutional Autoencoders), which achieve reasonable SNDR and R2 scores with minimal model size and MAC count.

    \item  We employ a novel hardware-aware balanced stochastic weight pruning scheme incorporating Linear Feedback Shift Registers (LFSRs) to reduce parameter memory requirements by up to 32.4\%.

    \item  We employ a memory optimization technique involving overlapping input and output activations on the same memory space, that reduces the peak activation memory storage by 37.5\%. The specific implementation details can be found in Section \ref{sec:RAMAN_features}.

\end{enumerate}

\par 
The rest of the paper is organized as follows: Section \ref{sec:proposed_scheme} introduces the proposed neural signal compression scheme utilizing CAE and explains the fundamentals of autoencoders. Section \ref{sec:RAMAN_architecture} describes the RAMAN architecture, its features, and the hardware-software co-optimization approach used to handle workload imbalances and minimize parameter memory requirements. This involves employing a hardware-aware balanced stochastic pruning scheme utilizing LFSRs. Section \ref{sec:results} presents the FPGA and ASIC implementation results, assesses model performance using SNDR and R2 score metrics across various CAE topologies, and compares the stochastic pruning scheme with the conventional magnitude-based pruning scheme. It also highlights the advantages of the stochastic pruning scheme in terms of memory size reduction. Finally, Section \ref{sec:conclusions} concludes the paper.


\section{Proposed compression scheme using convolutional autoencoders}
\label{sec:proposed_scheme}
The proposed neural signal compression scheme is depicted in Fig. \ref{fig:top_level_illust}. It comprises a signal conditioning module equipped with a neural amplifier and an analog-to-digital converter (ADC) for amplifying and digitizing the neural signal. Subsequently, the digitized neural signal is fed to the RAMAN tinyML accelerator for compression using convolutional autoencoders. All signal processing steps, including acquisition, amplification, digitization, and compression, occur in a head unit mounted on the head. The compressed data is transmitted wirelessly, while the decompression and decoding are conducted offline.
\subsection{Convolutional autoencoders}
\begin{figure}[h]
	\begin{center}
		\includegraphics[width = \columnwidth]{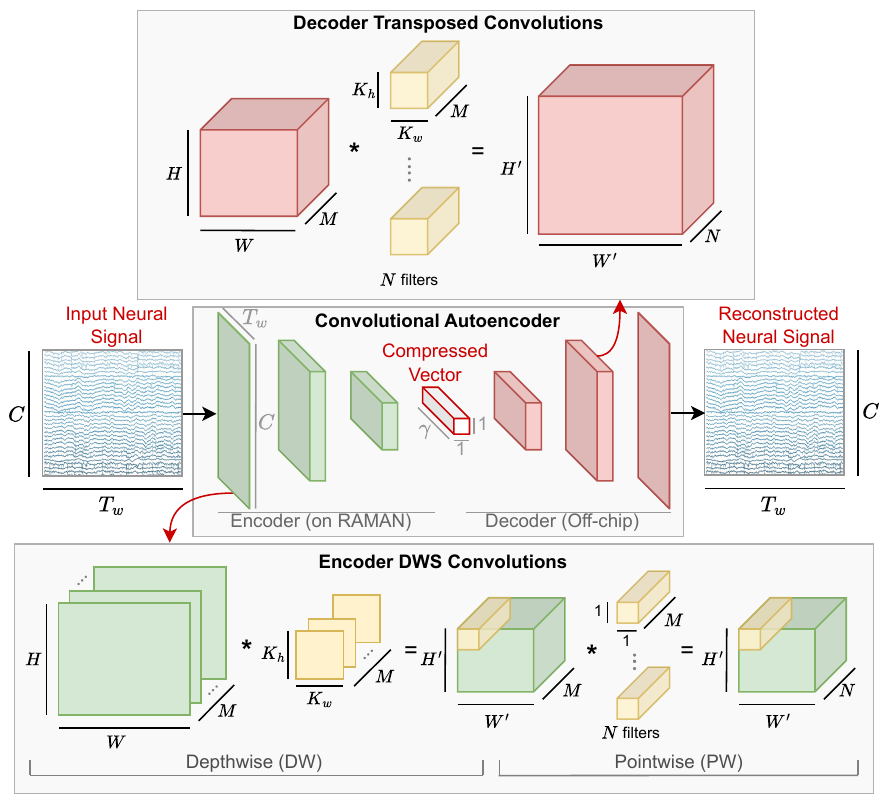}
	\end{center}
	\caption{A convolutional autoencoder with an encoder to convert the input neural signal window of size $C\times T_{w}$ to a compressed vector (in latent space) of size $1\times 1 \times \gamma$ and a decoder to recover back the original signal from the compressed vector. The encoder section employs depthwise separable (DWS) convolutions, while the decoder utilizes transposed convolutions.}
	\label{fig:cae}
\end{figure}
Convolutional autoencoders (CAEs) are a type of neural network architecture that combines the principles of convolutional neural networks (CNNs) with autoencoder structures. They are designed to learn efficient representations of input by encoding it into a lower-dimensional latent space and then reconstructing the original input from this representation as shown in Fig. \ref{fig:cae}. The input neural signal was divided into 50 ms windows at a 2 kHz sampling rate, giving 100 samples per window  ($T_{w}$). The CAE input is a 2D matrix sized  $C\times T_{w}$, where $C$ is the number of recording sites/channels set to 96 in our implementation. The encoder output is a vector of shape $1\times 1 \times \gamma$, representing the compressed data. Thus, the CR achieved by the proposed compression scheme is $C \times T_{w}/ \gamma.$
The CAEs are composed of two main parts: an encoder and a decoder. 
\subsubsection{Encoder/Compression}
The encoder part of a CAE typically consists of convolutional layers followed by pooling layers. These layers extract essential features from the input data and reduce its dimensionality. The encoding process can be represented as:
\begin{equation}
\mathbf{z} = f(\mathbf{W} * \mathbf{x} + \mathbf{b})
\end{equation}
where, \(\mathbf{x}\) is the input neural data, \(\mathbf{W}\) represents the convolutional filters, \(\mathbf{b}\) is the bias vector (typically with one value per output channel), \(\ast\) denotes the convolution operation, \(f\) is the activation function, such as ReLU (Rectified Linear Unit), applied element-wise and \(\mathbf{z}\) is the encoded representation (also called latent or compressed representation) of the input data. In our design, we employ depthwise separable (DWS) convolutions to minimize the number of operations. The DWS convolutions split the convolution process into a depth-wise and a point-wise convolution. The depthwise convolution operation can be represented as:
\begin{align*}
{\mathbf{OA}}_{h,w,m}^{l} =\  f\Biggl(& \mathbf{b}_{m}^{l\,(DW)} + 
 \sum\limits_{i=0}^{K_h-1} \sum\limits_{j=0}^{K_w-1} 
{(\mathbf{W}}_{i,j,m}^{l\,(DW)} \\ \cdot 
&\mathbf{IA}_{h\cdot s_h+i,w\cdot s_w+j,m}^{l}) \Biggr)
\end{align*}
where ${\mathbf{W}}^{l\,(DW)}$ is the depthwise convolutional kernel of size $K_h\times K_w \times M$ and $\mathbf{b}_m^{l\,(DW)}$ denotes the bias corresponding to the $m^{th}$ output channel for layer \lq$l$\rq . The $m^{th}$ filter in ${\mathbf{W}}^{l\,(DW)}$ is applied to the $m^{th}$ channel in the input activation ($\mathbf{IA}^{l}$) tensor to produce the $m^{th}$ channel of the filtered output activation ($\mathbf{OA}^{l}$) tensor. The strides along the height and width are denoted by $s_h$ and $s_w$, respectively, and the operator $\cdot$ represents scalar multiplication. The pointwise convolution operation can be expressed as:

\[
\mathbf{OA}_{h,w,n}^{l} =  f\Biggl( \mathbf{b}_{n}^{l\,(PW)} +  \sum\limits_{m=0}^{M-1} \mathbf{W}_{m,n}^{l\,(PW)} \cdot \mathbf{IA}_{h,w,m}^{l} \Biggl)
\]
where ${\mathbf{W}}^{l\,(PW)}$ is the pointwise convolutional kernel of size $M\times N$ and $\mathbf{b}_n^{l\,(PW)}$ denotes the bias corresponding to the $n^{th}$ output channel for layer \lq$l$\rq. $M$ is the number of input channels, and $N$ is the number of filters or output channels. 
Compared to the standard convolutions \cite{Krizhevsky2012}, the DWS convolutions reduce the computation by: 
\[\frac{1}{N} + \frac{1}{K_{h} \cdot K_{w}}\]
The input neural data $\mathbf{x}$ is provided to the first layer of the encoder as $\mathbf{IA}^0 = \mathbf{x}$. A series of convolutional and pooling operations is then applied, as described in Table~\ref{tab:model_arch}, to transform the input into a latent representation $\mathbf{z} = \mathbf{OA}^{L_{\text{enc}} - 1}$, where $L_{\text{enc}}$ is the total number of layers in the encoder. This latent vector $\mathbf{z}$ serves as the input to the decoder for signal reconstruction ($\hat{\mathbf{IA}}^0$ = z). The final output of the decoder, i.e., the reconstructed neural data, is denoted as $\hat{\mathbf{x}} = \hat{\mathbf{OA}}^{L_{\text{dec}} - 1}$, where $L_{\text{dec}}$ is the total number of decoder layers. The hidden layers of the encoder and decoder are indexed by $l = 1$ to $L_{\text{enc}} - 2$ and $l = 1$ to $L_{\text{dec}} - 2$, respectively. The output activations of a given layer serve as the input activations for the subsequent layer, expressed as $\mathbf{IA}^{l+1} = \mathbf{OA}^l$ for the encoder and $\hat{\mathbf{IA}}^{l+1} = \hat{\mathbf{OA}}^l$ for the decoder.

\subsubsection{Decoder/Decompression}
The decoder part of a CAE consists of upsampling or transposed convolutional layers. The purpose of the decoder is to reconstruct the original input data from the encoded representation. The decoding process can be represented as:
\begin{equation}
\mathbf{\hat{x}} = f(\mathbf{\hat{W}}*\mathbf{z} + \mathbf{\hat{b}})
\end{equation}
where, \(\mathbf{z}\) is the encoded representation obtained from the encoder, \(\mathbf{\hat{W}}\) represents the decoder filters, \(\mathbf{\hat{b}}\) is the bias vector, \(f\) is the activation function,  \(\mathbf{\hat{x}}\) is the reconstructed output data. We use transposed convolutions for decoding the compressed data, which can be represented as:
\begin{align*}
    \mathbf{\hat{OA}}_{h,w,n}^{l} = f\Biggl( & \mathbf{\hat{b}}_n^{l} + \sum\limits_{m=0}^{M-1} \sum\limits_{i=h_a}^{h_b} \sum\limits_{j=w_a}^{w_b} (\mathbf{\hat{W}}_{i,j,m,n}^{l} \\ \cdot & \mathbf{\hat{IA}}_{\frac{h-i}{s_h},\frac{w-j}{s_w},m}^{l}) \cdot \mathbbm{1}_Q \Biggl)
\end{align*}
where $h_a=\max\left(0, h-s_h(H-1)\right)$, $h_b=\min\left(h, K_h-1\right)$, $w_a=\max\left(0, w-s_w(W-1)\right)$, and $w_b=\min\left(w, K_w-1\right)$. $H$ and $W$ represent the height and width of the input activation tensor ($\mathbf{\hat{IA}}$), respectively. 
Additionally, $\mathbbm{1}_Q$, the indicator function of the statement $Q \equiv \left( \frac{h-i}{s_h} \in \mathbb{Z} \land \frac{w-j}{s_w} \in \mathbb{Z} \right)$, yields 1 if $Q$ is true, and 0 otherwise. Here $\mathbb{Z}$ denotes the set of integers.
Similarly, the depthwise transposed convolution operation can be expressed as:
\begin{align*}
    \mathbf{\hat{OA}}_{h,w,m}^{l} = f\Biggl(& \mathbf{\hat{b}}_m^{l\,(DW)} + \sum\limits_{i=h_a}^{h_b} \sum\limits_{j=w_a}^{w_b} (\mathbf{\hat{W}}_{i,j,m}^{l\,(DW)} \\ & \cdot \mathbf{\hat{IA}}_{\frac{h-i}{s_h},\frac{w-j}{s_w},m}^{l}) 
    \cdot \mathbbm{1}_Q \Biggl)
\end{align*}





where \(\mathbf{\hat{b}}^{l\,(DW)}\) and \(\mathbf{\hat{W}}^{l\,(DW)}\) denote the bias vector and kernel of the depthwise transposed convolution, respectively.
\par The objective of the CAE is to minimize the reconstruction error between the input data and the reconstructed output. A standard loss function used for this purpose is the Mean Absolute Error (MAE) loss, given by:
\begin{equation}
\mathcal{L} = \frac{1}{P} \sum_{i=1}^{P} \left|\mathbf{x}_i - \mathbf{\hat{x}}_i\right|
\end{equation}
where \(P\) is the number of data samples, \(\mathbf{x}_i\) and \(\mathbf{\hat{x}}_i\) are the \(i^{th}\) input and reconstructed output data samples, respectively.
By minimizing this loss function, the CAE learns to encode the input data into a lower-dimensional representation while retaining important information for accurate reconstruction.

\section{Hardware Architecture}
\label{sec:RAMAN_architecture}
This section introduces RAMAN architecture, features, and hardware-aware balanced stochastic pruning scheme. 
\subsection{Top-Level Architecture} 
\begin{figure}[h]
	\begin{center}
		\includegraphics[width = \columnwidth]{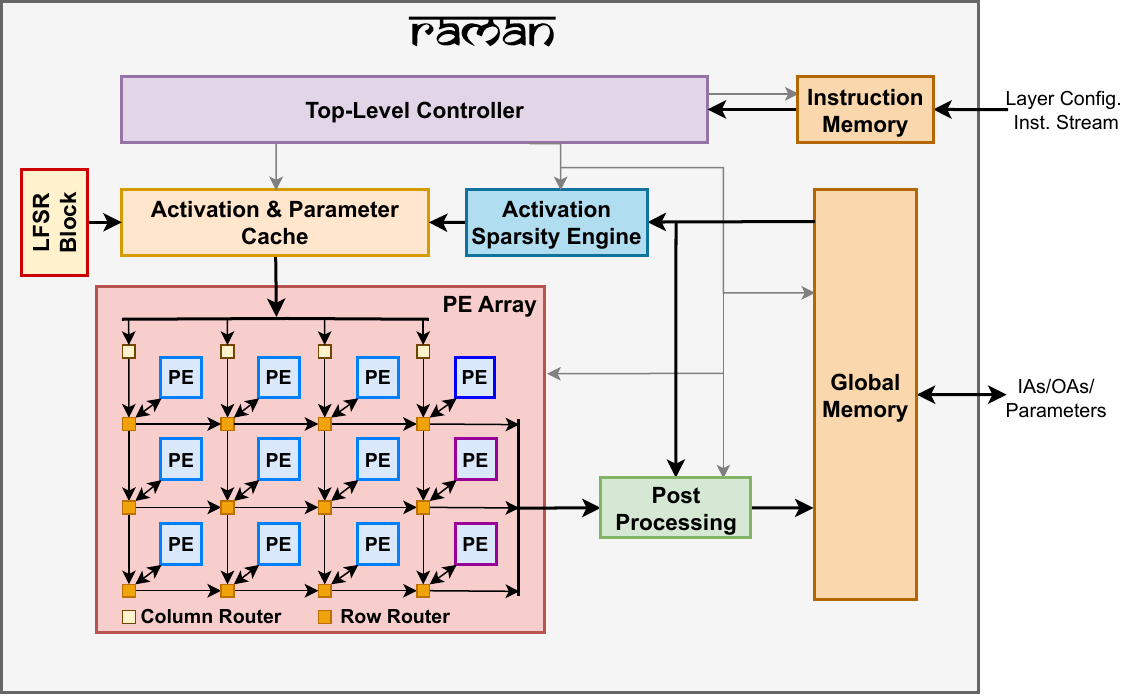}
	\end{center}
	\caption{Top-Level architecture of the RAMAN accelerator.} 
	\label{fig:top_module}
\end{figure}

The top-level architecture of the RAMAN accelerator utilized to deploy the encoder of CAE for neural data compression is depicted in Fig. \ref{fig:top_module}. The architecture comprises compute, memory, and control subsystems. The computing subsystem comprises a processing element (PE) array for performing MAC operations, an activation sparsity engine to exploit sparsity in activations, and a post-processing module (PPM) responsible for the rectified linear unit (ReLU) activation, quantization, pooling, bias addition, and residual addition. The memory subsystem consists of a global memory to store activations and parameters and a cache to exploit temporal data reuse. The control subsystem includes a top-level controller that schedules and sequences different operations and issues commands to various processing and memory blocks. A comprehensive description of the architecture is presented in \cite{RAMAN2023}. In this paper, we introduce a novel stochastic pruning scheme for weight pruning, that eliminates the indexing overhead of the compressed weights compared to our previous implementation that utilizes conventional magnitude-based pruning\cite{RAMAN2023,Krishna2024biocas,Krishna2023,Krishna2023PrimeAsia,KrishnaISCAS2024}. 

\subsection{RAMAN features}
    \begin{figure}[h!]
	\begin{center}
		\includegraphics[width = \columnwidth]{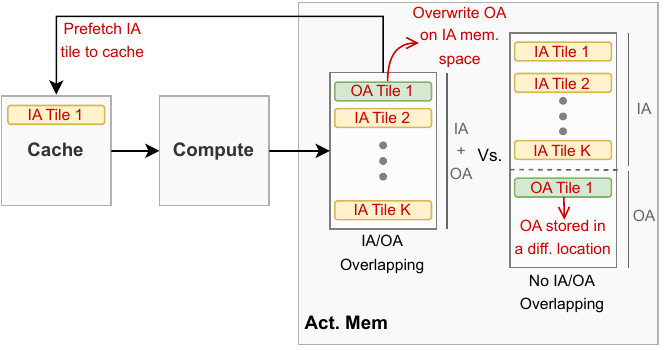}
	\end{center}
        \caption{Peak activation memory reduction using IA/OA memory overlapping.}
	\label{fig:ia_oa_overlap}
    \end{figure}
\label{sec:RAMAN_features}

    \begin{figure*}[t!]
	\begin{center}
		\includegraphics[width = 18cm]{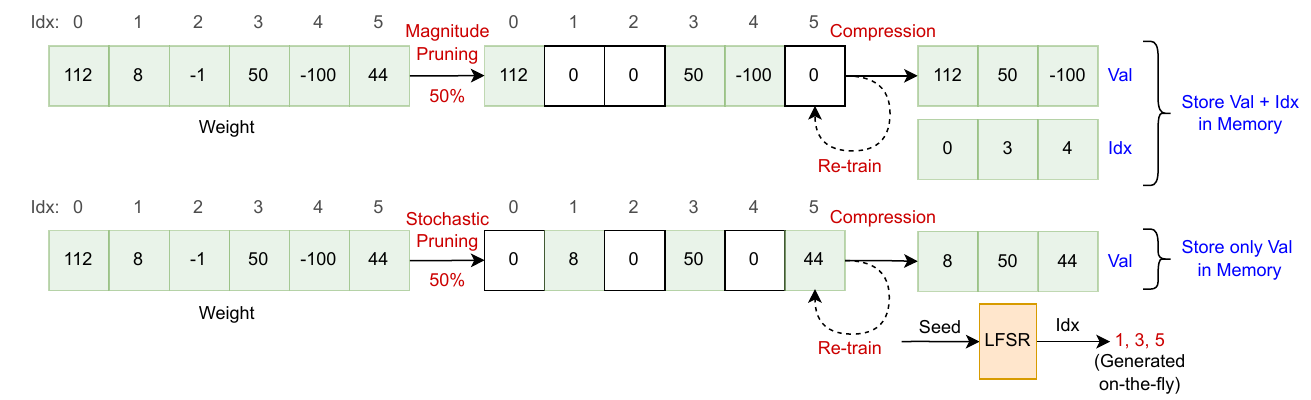}
	\end{center}
	\caption{Illustration of stochastic pruning compared with conventional magnitude-based pruning. The magnitude-based pruning retains the stronger weights and prunes the rest. The sparse weight vector is then compressed and stored as a value (Val) and index (Idx) pair. On the other hand, stochastic pruning retains the weight values whose indices are covered by LFSR-generated pseudo-random sequence. Since the indices are generated on the fly during inference, only the weight values are stored in the memory, eliminating the need for explicit index storage. The LFSR's seed (initial state of flip flops) and structure (feedback polynomial) are kept constant during training and inference.} 
	\label{fig:stochastic_w_pruning}
\end{figure*}
\begin{figure}[h!]
	\begin{center}
		\includegraphics[width = \columnwidth]{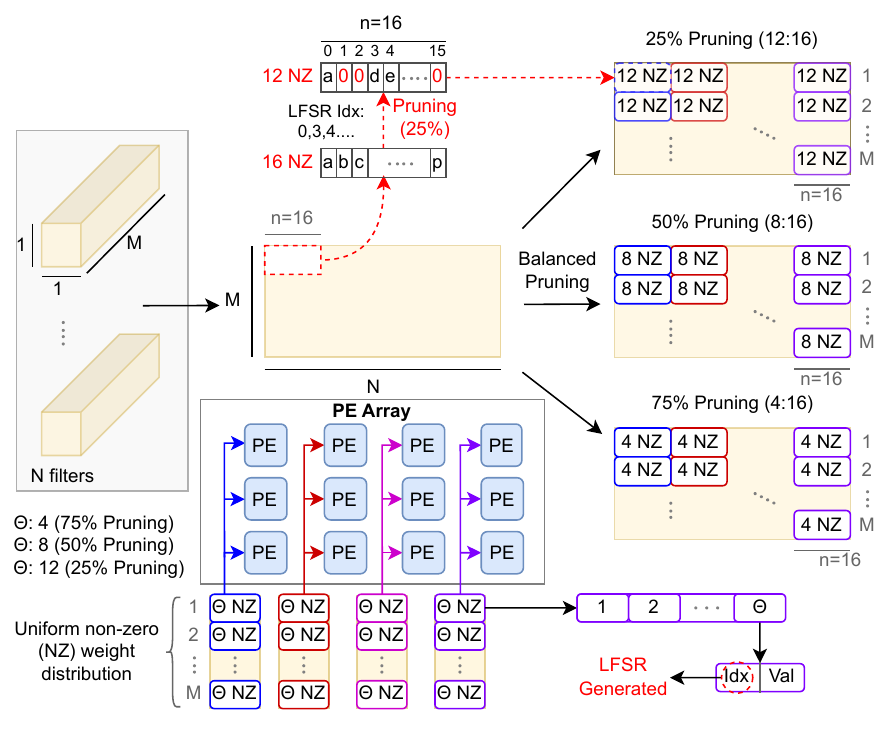}
	\end{center}
	\caption{The balanced pruning strategy for different pruning percentages. Depending on the pruning percentage, each PE receives an equal number of non-zero weight elements, leading to a uniform workload across PEs. The number of non-zero weight elements ($\Theta$) in a given tile after pruning is fixed for a given pruning percentage set at 4, 8, and 12 for 75\%, 50\%, and 25\% pruning, respectively. The indices of the compressed weights are generated by the LFSR and not stored explicitly in memory.} 
	\label{fig:habsp}
\end{figure}

The architectural features of the RAMAN tinyML accelerator tailored for edge computing are outlined as follows:
\begin{enumerate}
    \item Sparsity: RAMAN exploits both input activation and weight sparsity to reduce latency, memory access, and storage. RAMAN can skip the processing cycles with zero activations and weights to minimize the processing latency. The sparse weight matrices are compressed and stored in the memory. The stochastic pruning scheme presented in Section \ref{sec:HA-BSP} generates the compressed weight indices on the fly, eliminating the need for explicit index storage in memory. 
    \item Dataflow: RAMAN employs Gustavson inspired \cite{Zhang2021Gamma} dataflow with optimal input and output activation reuse to reduce memory access. The dataflow reduces the partial sums (Psums) within the processing element to eliminate the Psum writeback traffic. 
    \item Peak Activation Memory Reduction: The state-of-the-art accelerators typically segment the Input Activations (IAs) and Output Activations (OAs) within the memory, with the total activation memory being the sum of IA and OA memory spaces. In our approach, we eliminate logical partitioning, enabling OAs to directly overwrite the IA memory space, as illustrated in Figure \ref{fig:ia_oa_overlap}. This is achieved by pre-fetching the IA tile into the cache, making the original content in the activation memory redundant, and enabling OA overwriting within the same memory space. A detailed explanation can be found in \cite{RAMAN2023}.

    \item Programmability: RAMAN is capable of supporting various types of neural network topologies, including standard CNN models as well as separable convolutions. The separable convolution models comprise depth-wise (DW) and point-wise (PW) layers, which effectively reduce MAC operations compared to standard convolutions \cite{Howard2017}. Additionally, RAMAN can handle operations such as max pooling, average pooling, and fully connected (FC) layers.
\end{enumerate}

\subsection{Hardware-aware Balanced Stochastic Pruning} 
\label{sec:HA-BSP}

We present a novel hardware-aware balanced stochastic weight pruning scheme aimed at reducing memory storage, access, and latency. Traditionally, magnitude-based pruning removes weaker synaptic weights while retaining stronger ones. However, this approach involves storing both indices and non-zero weight values, leading to additional indexing overhead as shown in Fig. \ref{fig:stochastic_w_pruning}. Using magnitude-based pruning, our previous implementation \cite{Krishna2024biocas,RAMAN2023} utilized 8-bit weights and 4-bit indices to store compressed non-zero weights. In this work, we introduce a stochastic pruning approach that eliminates index storage in memory, storing only the non-zero weight values. Karimzadeh et al. \cite{Karimzadeh2021} propose a similar pruning strategy based on the LFSR-generated pseudo-random sequences, primarily focusing on software implementation and analysis. However, their study does not extend to exploring the hardware feasibility of integrating this strategy into an actual ML accelerator.

\par 
During the training phase, this scheme generates a pseudo-random sequence (PRS) serving as indices to sparsify the weight matrix. The synaptic weights corresponding to the indices covered by the PRS are retained, while the remaining weights are forced to zero. Subsequently, the un-pruned weight values (covered by the PRS) are retrained in the subsequent step.
The pseudo-random sequence is generated using a linear feedback shift register (LFSR), offering a straightforward implementation during inference. This method enables the on-the-fly generation of indices without the need to store them in memory, thereby minimizing storage overhead. Additionally, implementing the logic for the LFSR incurs only a minimal increase in area and resource utilization, as demonstrated in Section \ref{sec:result_FPGA_imp}.
\par

During deployment, only the non-zero weight values are stored in the RAMAN parameter memory. The same seed and LFSR structure employed during training are used to generate an identical PRS sequence during inference. The non-zero weight values are then read from memory and matched with the indices generated by the LFSR. This matching ensures that each non-zero weight value is correctly multiplied by its corresponding activation value. Ultimately, the indices generated by the LFSR guarantee that all non-zero weights stored in the RAMAN memory are covered, facilitating the computation of the final output. 
\par

Additionally, the balanced pruning strategy ensures a uniform distribution of non-zero weights across weight tiles, effectively mitigating workload imbalances. Without this strategy, non-zero weights would be unevenly distributed among various weight tiles processed by different processing elements (PEs), resulting in workload discrepancies and performance limitations imposed by the PE handling the heaviest load. The operation of the balanced pruning strategy is illustrated in Fig. \ref{fig:habsp}. Initially, N filters with an input channel dimension of M are represented as a weight matrix of size $M \times N$. Subsequently, the weight matrix is divided into tiles of size $1 \times n$, where $n$ is determined by the depth of the PE register-file (RF), set to 16 in our design. Each $1 \times n$ tile is then pruned based on the required sparsity level to have $\Theta$ non-zeros, where $\Theta$ is a function of the pruning percentage. For example, for 25\%, 50\%, and 75\% pruning percentages, $\Theta$ is set to  12, 8, and 4, respectively. The weight values with indices not covered by LFSR-generated PRS are pruned. Furthermore, the LFSR-generated indices ensure that the same number of weights are pruned in each tile, resulting in structured sparsity that can be efficiently leveraged in RAMAN. Following balanced pruning, each tile contains an equal number of non-zero weights processed by different PE columns, achieving uniform workload distribution across the PE array. Each processing element (PE) contains four multiply-accumulate (MAC) units, executing four MAC operations in a single cycle and necessitating four non-zero weights per cycle. Thus, we employ four 4-bit LFSRs to generate four random indices simultaneously. With 4 MACs per PE, processing each weight tile requires $\Theta/4$ cycles and, consequently, $\Theta/4$ cycles to generate $\Theta$ indices using the 4 LFSRs. The initial seed of each LFSR is selected to ensure that unique $\Theta$ indices are generated for each $1 \times n$ weight tile. This approach prevents repeated indices within a tile, thereby ensuring the desired $\Theta$ non-zero elements are produced. The detailed PE architecture and operation are presented in \cite{RAMAN2023}.
\par Using LFSRs for generating PRS offers several advantages: 1) LFSRs can be implemented using a small number of flip-flops and XOR gates, making them highly compact in terms of hardware resources. 2) LFSRs generate pseudo-random sequences on-the-fly without any memory footprint. 3) The PRS generated by LFSRs are deterministic and repeatable, meaning that the same initial state (seed) and the feedback polynomial will always produce the same sequence. This property is desirable for our use case to ensure the same PRS generation for both training and inference. 4) LFSRs typically consume less power compared to some other methods of generating pseudo-random sequences, especially when implemented in hardware. Moreover, in our scenario, the need for costly additional memory access to retrieve indices is eliminated as LFSR logic replaces index memory storage.

\section{Results}
In this section, we present the FPGA and ASIC implementation results of RAMAN and evaluate the performance of CAE models across various topologies and sparsity levels. We compare the stochastic pruning scheme with the standard magnitude pruning scheme. Furthermore, we compare our neural signal compression scheme employing CAEs with existing literature.

\vspace{-1em}

\label{sec:results}
\subsection{FPGA Implementation}
\label{sec:result_FPGA_imp}

The RAMAN accelerator was implemented on Efinix Titanium Ti60 FPGA, incorporating the stochastic weight pruning scheme presented in Section \ref{sec:HA-BSP}. The hardware specifications of the RAMAN architecture and resource utilization details are presented in Table \ref{tab:raman_specs} evaluated for a custom DS-CAE1 model and the MobileNetV1-based autoencoder model with a width multiplier of 0.25x, represented as MobileNetV1-CAE(0.25x). The specific MobileNetV1-CAE and DS-CAE model architectures are provided in Table \ref{tab:mob_1_table} and \ref{tab:ds_cae_table}. 
\par 
The encoder model of the CAE utilizes CONV, DW, PW, and pooling layers, amounting to a total of 2.234 million MAC operations for the DS-CAE1 model and 22.91 million MACs for the MobileNetV1-CAE(0.25x) model. The PW layer accounts for the majority of these MAC operations. RAMAN employs 12 processing elements, with each PE containing four MAC units. The register-file memory utilization is 0.896 kB. The PE array utilizes 52\% (0.576 kB) of the registers, as each PE consists of a 16 × 24b Register File (RF) to store the Psums. Additionally, RAMAN offers the flexibility to decrease the PE RF width to 20b or even lower, depending on the application, thereby reducing register utilization. The post-processing module (PPM) stores post-processing parameters in registers, resulting in 32\% (0.32 kB) register utilization.
\begin{table}[h]
\centering
\renewcommand{\arraystretch}{1.2}
\caption{Specifications and FPGA resource utilization.}
\begin{tabular}{|c|c|c|c|c|c|c|c|}
\hline
\textbf{Platform} &  Efinix Ti60\\ \hline
\textbf{Layers Supported} &  \begin{tabular}[c]{@{}c@{}} CONV, DW, PW, \\   FC and Max/Average pooling.   \end{tabular} \\ \hline
\textbf{Number of PEs} & 12 (4 MACs/PE) \\ \hline
\textbf{Reg-file Memory} &   \begin{tabular}[c]{@{}c@{}} 0.896 kB   \end{tabular}  \\ \hline
\textbf{Clock Rate} & 2-7  MHz \\ \hline
\textbf{Precision}  & \begin{tabular}[c]{@{}c@{}} Weights \& Activations: 8b fixed point, \\  Partial-sums: 24b fixed point.   \end{tabular}  \\ \hline
\textbf{Power (mW)} &  \begin{tabular}[c]{@{}c@{}} 47.91$^\dagger$ (Dynamic: 3.11, Static: 44.79) @ 2 MHz \\ 53.97$^\star$ (Dynamic: 8.97, Static: 45) @ 7 MHz    \end{tabular}    \\ \hline
\textbf{XLR cells}  & \begin{tabular}[c]{@{}c@{}} 52.3k (86\% util.), 37.3k LUTs \& 8.6k FFs    \\  LFSR logic: 46 XLR cells (32 LUTs \& 20 FFs)  \end{tabular} \\  \hline
\textbf{DSPs} & 61 (38.12\% util.) \\ \hline
\textbf{\begin{tabular}[c]{@{}c@{}} Memory Blocks \\ (10 kB Blocks)  \end{tabular}} &  \begin{tabular}[c]{@{}c@{}} 92$^\dagger$ (Param.: 10, Act.: 48, Cache: 27, Rest: 7) \\ 151$^\star$(Param.: 69, Act.: 48, Cache: 27, Rest: 7) \end{tabular}   \\ \hline 
\textbf{\begin{tabular}[c]{@{}c@{}}  MAC Operations \\ (in Millions)  \end{tabular}} & \begin{tabular}[c]{@{}c@{}} \begin{tabular}[c]{@{}c@{}} 2.234$^\dagger$ (CONV: 15.47\% , DW: 12.92\%, \\ PW: 71.22\%, Pool: 0.39\%)  \end{tabular} \\  \begin{tabular}[c]{@{}c@{}} 22.91$^\star$  (CONV: 1.51\% , DW: 8.18\%, \\ PW: 90.29\%, Pool: 0.02\%)  \end{tabular} \end{tabular}   \\ \hline
\textbf{Latency (ms)} &  45.47$^\dagger$  @ 2 MHz, 47.82$^\star$ @ 7 MHz\\ \hline
\textbf{ \begin{tabular}[c]{@{}c@{}} Parameter  \\  Memory (kB)\end{tabular}} & \begin{tabular}[c]{@{}c@{}} \begin{tabular}[c]{@{}c@{}} Baseline floating point: 45.76$^\dagger$, 8b Quantized +\\   75\% PW Stochastic Pruning: 6.19$^\dagger$ \end{tabular} \\  \begin{tabular}[c]{@{}c@{}} Baseline floating point: 841.92$^\star$, 8b Quantized +\\   75\% PW Stochastic Pruning: 76.08$^\star$ \end{tabular} \end{tabular}  \\ \hline
\end{tabular}
\label{tab:raman_specs}
\begin{tablenotes}
      \small
       \item XLR: eXchangeable Logic and Routing (XLR) cell, DSP: Digital Signal Processor
       \item $^\dagger$DS-CAE1 model, $^\star$MobileNetV1-CAE(0.25x) model
    \end{tablenotes}
\vspace*{-0.1in}  
\end{table}
\par 
The power consumption is estimated to be 47.91 mW (3.11 mW dynamic power + 44.79 mW static power) at a 2 MHz clock for the DS-CAE1 model, and 53.97 mW (8.97 mW dynamic power + 45 mW static power) at a 7 MHz clock for the MobileNetV1-CAE(0.25x) model. Notably, owing to the FPGA implementation, the static power dissipation is significant. 
The architecture utilizes 37.3k 4-input look-up tables (LUTs) and 8.6k flip-flops (FFs), which are mapped as eXchangeable Logic and Routing (XLR) cells on the Efinix FPGA \cite{EfinixTi60}. Table \ref{tab:raman_specs} illustrates that the logic overhead associated with the LFSR for enabling stochastic pruning is minimal with 32 LUTs and 20 FFs totaling 46 XLR cells. 
\par 
The activations and weights of the encoder network were quantized to 8 bits, while the Psums generated after the MAC operation were represented with 24 bits. Consequently, the parameter memory requirement decreased from 45.76 kB to 6.19 kB (7.4x reduction) and from 841.92 kB to 76.08 kB (11x reduction) after 8-bit quantization and 75\% point-wise stochastic weight pruning for the DS-CAE1 and MobileNetV1-CAE(0.25x) models, respectively. In addition, the IA and OA memory overlapping scheme outlined in \cite{RAMAN2023} reduced the peak activation memory of the MobileNetV1-CAE(0.25x) model from 76.8 kB to 48 kB, representing a 37.5\% reduction. The global memory, comprising parameter memory and activation memory, along with the cache, were mapped to 10 kB block memory units available in the Efinix Ti60 FPGA \cite{EfinixTi60}. The latency of RAMAN for a single input inference to generate the compressed representation is measured at 45.47 ms at a clock rate of 2 MHz and 47.82 ms at a clock rate of 7 MHz for the DS-CAE1 and MobileNetV1-CAE(0.25x) models, respectively. Since the input window spans 50 ms, the achieved latency can effectively process the input within a given period. The achieved throughput of the RAMAN accelerator is 98.3 MOp/s for the DS-CAE1 model at 2 MHz and 958.3 MOp/s for the MobileNetV1-CAE(0.25x) model at 7 MHz.

\subsection{ASIC Implementation}
\label{sec:result_ASIC_imp}
\begin{figure}[h!]
	\begin{center}
		\includegraphics[width = 7cm]{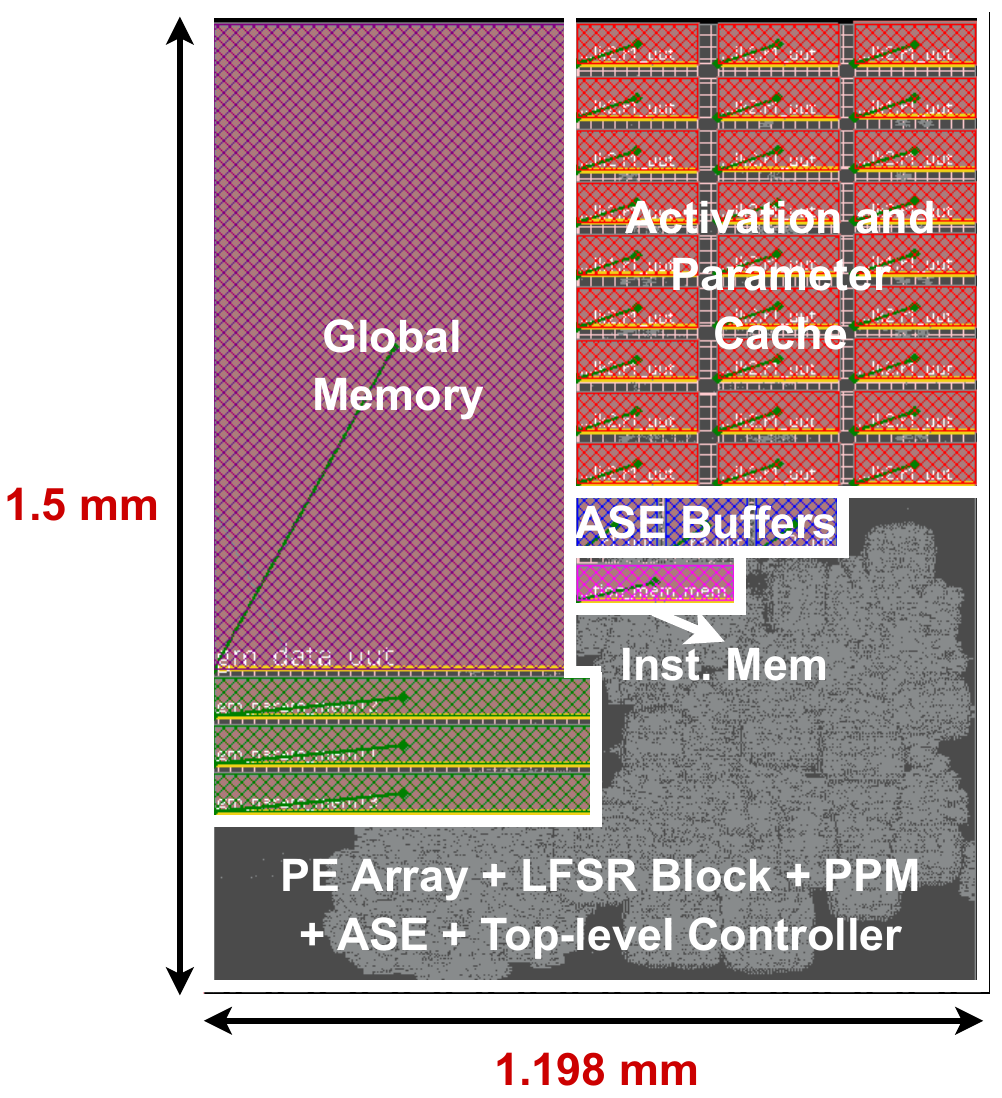}
	\end{center}
	\caption{Layout of the RAMAN encoder in TSMC 65-nm CMOS process technology, occupying  1.8 mm$^2$.} 
	\label{fig:RAMAN_CAE_BCI_Layout}
\end{figure}
The RAMAN encoder was designed and simulated using a TSMC 65-nm CMOS process technology, with a core area of 1.8 mm$^2$, which corresponds to 0.0187 mm$^2$ per channel. The design integrates a total of 63.36 kB of on-chip SRAM memory. Since RAMAN targets energy-constrained edge computing applications, it relies exclusively on on-chip SRAM for storing model parameters and activations. The layout of the RAMAN encoder, generated through full back-end flow, is shown in Fig. \ref{fig:RAMAN_CAE_BCI_Layout}. All results, including area, timing, and power estimates, were obtained through a complete ASIC design flow, utilizing Cadence Genus for logic synthesis, Cadence Innovus for place-and-route, and Cadence Voltus for power analysis. The estimated power consumption is 1.45 mW, which corresponds to 15.1 $\mu$W per channel at a clock frequency of 2 MHz and a supply voltage of 1.2 V for the DS-CAE1 model.

\subsection{Dataset Description}
We utilized the dataset from \cite{GnodeDataset}, comprising neural recordings from the motor cortex of two macaque monkeys, K and L, during an instructed reach and grasp task. Neural activity was captured using 10-by-10 Utah electrode arrays with only 96 active electrodes. Each recording session includes two recordings, one for each monkey. The signals were sampled at 30 kHz. 
\par
In this work, we are interested in the local field potential (LFP) signals, which are the aggregate synaptic activities of populations of neurons. Given that the frequencies of interest in LFPs typically range from 0.1 to 300 Hz, the sampling rate can be substantially reduced compared to what is necessary for spike-based processing, which typically falls between 1 and 2 kS/s. Thus, the signals were downsampled to 2 kS/s after applying a low-pass FIR filter of order 20 with a cut-off frequency of 1 kHz. The training, validation, and test sets were split into time windows of 50 ms, each corresponding to 100 samples per window at a sampling rate of 2 kHz. Input to the CAE is a 2D matrix of size 96 (Channels) x100 (samples per window). From each recording session, the first 80\% of the recording is used for training, the next 10\% for validation, and the final 10\% for testing.
The training is conducted offline and the entire implementation pipeline can be divided into three phases:
\begin{enumerate} 
\item Training phase: During this phase, the RAMAN encoder is disabled and the uncompressed data is transmitted for offline training. Since real-time processing is not necessary, data can be recorded in local memory (such as an SD card) and transmitted at a lower data rate (channel-wise). The entire encoder and decoder stack of the CAE is trained offline, and the encoder parameters after training are stored in on-chip RAMAN memory. Additionally, the encoder model topology is encoded and stored as instructions in the instruction memory for programming RAMAN. The training phase occurs only initially. 
\item Deployment phase: During this phase, compressed data from RAMAN is transmitted for real-time offline decoding or analysis. The encoder model parameters and instructions stored in RAMAN are used for real-time data encoding (compression).
\item Calibration phase: During this phase, the model is calibrated to compensate for electrode drifts and other non-idealities over time. The amount of data required for calibration is significantly smaller than that needed for re-training the model \cite{Valencia2024}. Since the data size is small, it can be transmitted at a lower bandwidth and does not need to be in real-time. This phase can occur regularly for the periodic calibration of the model, and it has been reported that model calibration achieves similar performance to complete re-training \cite{Valencia2024}. RAMAN encoder parameters remain static, and only the decoder is trained offline eliminating the need for training within the head unit. Several factors influence the calibration frequency, including the type of electrode, the implantation site, and the specifics of the experimental conditions. The model's performance can be empirically assessed for calibration by periodically transmitting the raw signal and comparing the reconstruction error between the original raw signal and the reconstructed signal from the decoder.
\end{enumerate}

\begin{table}[h]
\caption{Architecture of the models.}
\label{tab:model_arch}
\begin{subtable}{\columnwidth}
\centering
\begin{tblr}{
  cell{2}{1} = {r=11}{},
  cell{13}{1} = {r=11}{},
  cell{1-8,10-15,17-23}{2} = {c=2}{},
  vline{2, 4-6} = {-}{},
  hlines = {-}{},
  rows = {rowsep=1pt},
  columns = {colsep=4pt},
}
\hline
Stage   & Type / Stride &   & M & N & Output Size   \\
\begin{sideways}Encoder\end{sideways}   & Conv (3$\times$3) / s2    &   & 1 & 32    & 48$\times$50$\times$32    \\
& Conv dws (3$\times$3) / s1$^\ddagger$    &           & 32    & 64    & 48$\times$50$\times$64    \\
& Conv dws (3$\times$3) / s2    &           & 64    & 128   & 24$\times$25$\times$128   \\
& Conv dws (3$\times$3) / s1    &           & 128   & 128   & 24$\times$25$\times$128   \\
& Conv dws (3$\times$3) / s2    &           & 128   & 256   & 12$\times$13$\times$256   \\
& Conv dws (3$\times$3) / s1    &           & 256   & 256   & 12$\times$13$\times$256   \\
& Conv dws (3$\times$3) / s1    &           & 256   & 512   & 12$\times$13$\times$512   \\
& 5$\times$ & Conv dws (3$\times$3) / s1    & 512   & 512   & 12$\times$13$\times$512   \\
& Conv dws (3$\times$3) / s2    &           & 512   & 1024  & 6$\times$7$\times$1024    \\
& Conv dws (3$\times$3) / s1    &           & 1024  & 1024  & 6$\times$7$\times$1024    \\
& Avg Pool (6$\times$7) / s1    &           & 1024  & 1024  & \fbox{1$\times$1$\times$1024$^\dagger$}   \\
\begin{sideways}Decoder\end{sideways}   & ConvTranspose dw (6$\times$7) / s1$^\S$    &   & 1024  & 1024  & 6$\times$7$\times$1024    \\
& ConvTranspose (3$\times$3) / s1$^\P$   &           & 1024  & 1024  & 6$\times$7$\times$1024    \\
& ConvTranspose (3$\times$3) / s2   &           & 1024  & 512   & 12$\times$13$\times$512   \\
& 5$\times$ & ConvTranspose (3$\times$3) / s1   & 512   & 512   & 12$\times$13$\times$512  \\
& ConvTranspose (3$\times$3) / s1   &           & 512   & 256   & 12$\times$13$\times$256   \\
& ConvTranspose (3$\times$3) / s1   &           & 256   & 256   & 12$\times$13$\times$256   \\
& ConvTranspose (3$\times$3) / s2   &           & 256   & 128   & 24$\times$25$\times$128   \\
& ConvTranspose (3$\times$3) / s1   &           & 128   & 128   & 24$\times$25$\times$128   \\
& ConvTranspose (3$\times$3) / s2   &           & 128   & 64    & 48$\times$50$\times$64    \\
& ConvTranspose (3$\times$3) / s1   &           & 64    & 32    & 48$\times$50$\times$32    \\
& ConvTranspose (3$\times$3) / s2   &           & 32    & 1     & 96$\times$100$\times$1
\end{tblr}
\caption{ MobileNetV1-CAE(1x) model.}
\label{tab:mob_1_table}
\end{subtable}
\begin{subtable}{\columnwidth}
\centering
\begin{tblr}{
  cell{2}{1} = {r=5}{},
  cell{7}{1} = {r=5}{},
  cell{1-4,6-7,9-11}{2} = {c=2}{},
  vline{2, 4-6} = {-}{},
  hlines = {-}{},
  rows = {rowsep=1pt},
  columns = {colsep=4pt},
}
\hline
Stage   & Type / Stride &   & M & N & Output Size   \\
\begin{sideways}Encoder\end{sideways}   & Conv (3$\times$3) / s2    &   & 1 & 16    & 48$\times$50$\times$16    \\
& Conv dws (3$\times$3) / s2$^\ddagger$    &           & 16    & 16    & 24$\times$25$\times$16    \\
& Conv dws (3$\times$3) / s2    &           & 16    & 64    & 12$\times$13$\times$64    \\
& n$^{\star}\times$ & Conv dws (3$\times$3) / s1    & 64    & 64    & 12$\times$13$\times$64   \\
& Avg Pool (12$\times$13) / s1    &           & 64    & 64    & \fbox{1$\times$1$\times$64$^\dagger$}   \\
\begin{sideways}Decoder\end{sideways}   & ConvTranspose dw (12$\times$13) / s1$^\S$    &   & 64    & 64    & 12$\times$13$\times$64    \\
& n$^{\star}\times$ & ConvTranspose (3$\times$3) / s1$^\P$   & 64   & 64   & 12$\times$13$\times$64  \\
& ConvTranspose (3$\times$3) / s2   &           & 64   & 16   & 24$\times$25$\times$16   \\
& ConvTranspose (3$\times$3) / s2   &           & 16   & 16    & 48$\times$50$\times$16    \\
& ConvTranspose (3$\times$3) / s2   &           & 16    & 1     & 96$\times$100$\times$1
\end{tblr}
\caption{DS-CAE models.}
\label{tab:ds_cae_table}
\end{subtable}
\begin{tablenotes}
  \small
  \item $^\ddagger$ A `Conv dws (3$\times$3)' layer consists of a Depthwise convolutional layer with filter shape 3$\times$3$\times$M, followed by a Pointwise convolutional layer with filter shape 1$\times$1$\times$M$\times$N.
  \item $^{\dagger}$ Compressed representation of the input signal.
  \item $^\S$ The filter shape of a `ConvTranspose dw ($K_h\times K_w$)' layer is $K_h \times K_w \times$M.
  \item $^\P$ The filter shape of a `ConvTranspose (3$\times$3)' layer is 3$\times$3$\times$M$\times$N.
  \item $^{\star}$ $\text{n}=2$ for DS-CAE1 and $\text{n}=1$ for DS-CAE2.
\end{tablenotes}
\end{table}

\begin{figure*}[t]
    \centering
    \begin{subfigure}[]{0.49\textwidth}
        \centering
        \includegraphics[width=\textwidth]{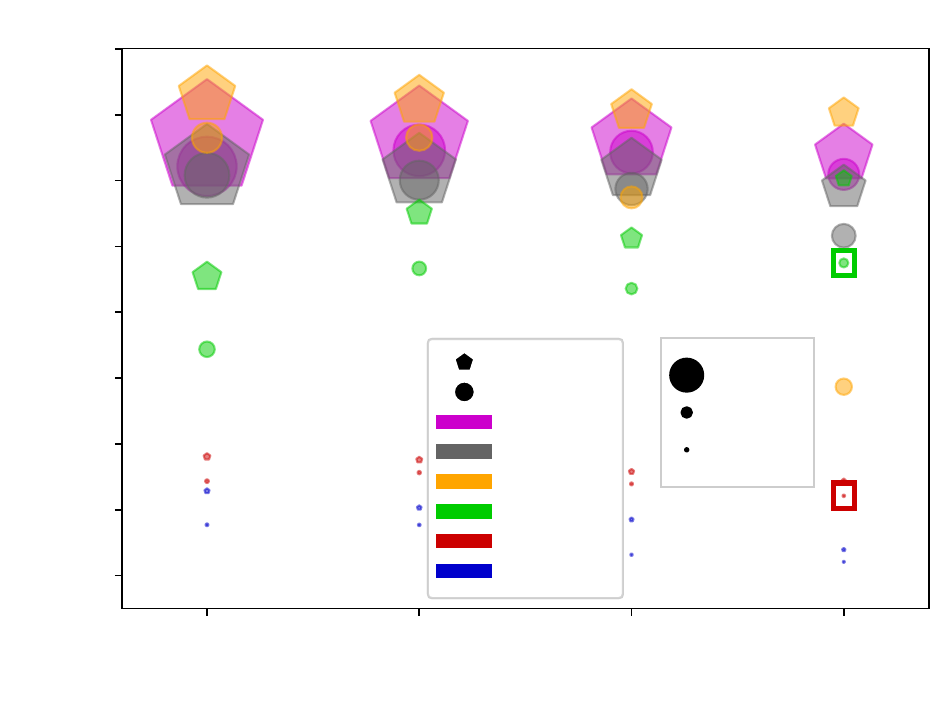}
        \caption{}
        \label{fig:Ablation_a}
    \end{subfigure}
    ~
    \begin{subfigure}[]{0.49\textwidth}
        \centering
        \includegraphics[width=\textwidth]{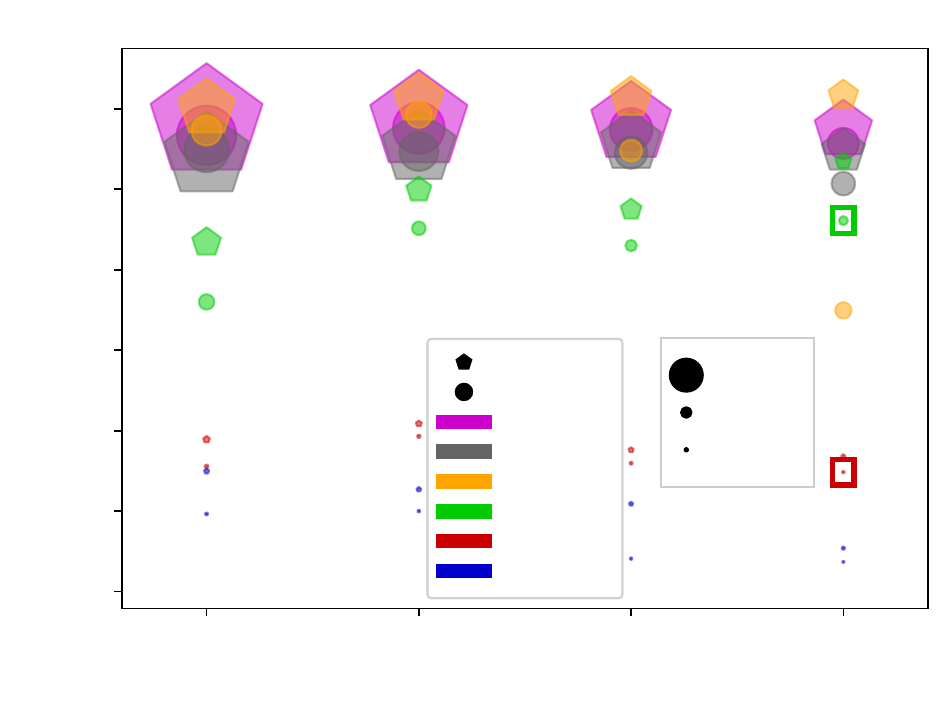}
        \caption{}
        \label{fig:Ablation_b}
    \end{subfigure}
    \\
    \begin{subfigure}[]{0.49\textwidth}
        \centering
        \includegraphics[width=\textwidth]{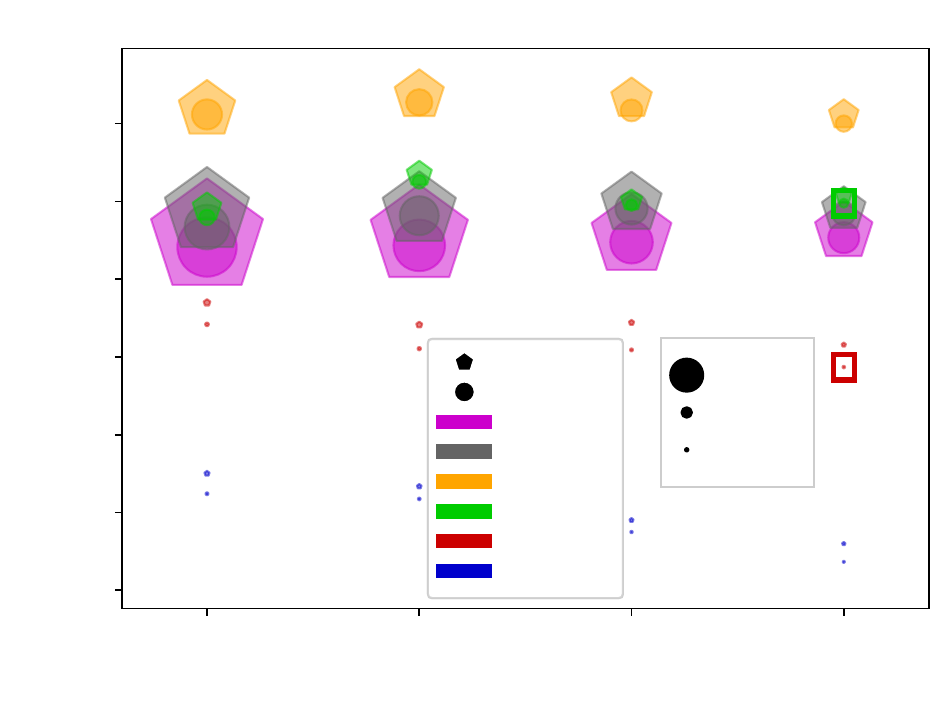}
        \caption{}
        \label{fig:Ablation_c}
    \end{subfigure}
    ~
    \begin{subfigure}[]{0.49\textwidth}
        \centering
        \includegraphics[width=\textwidth]{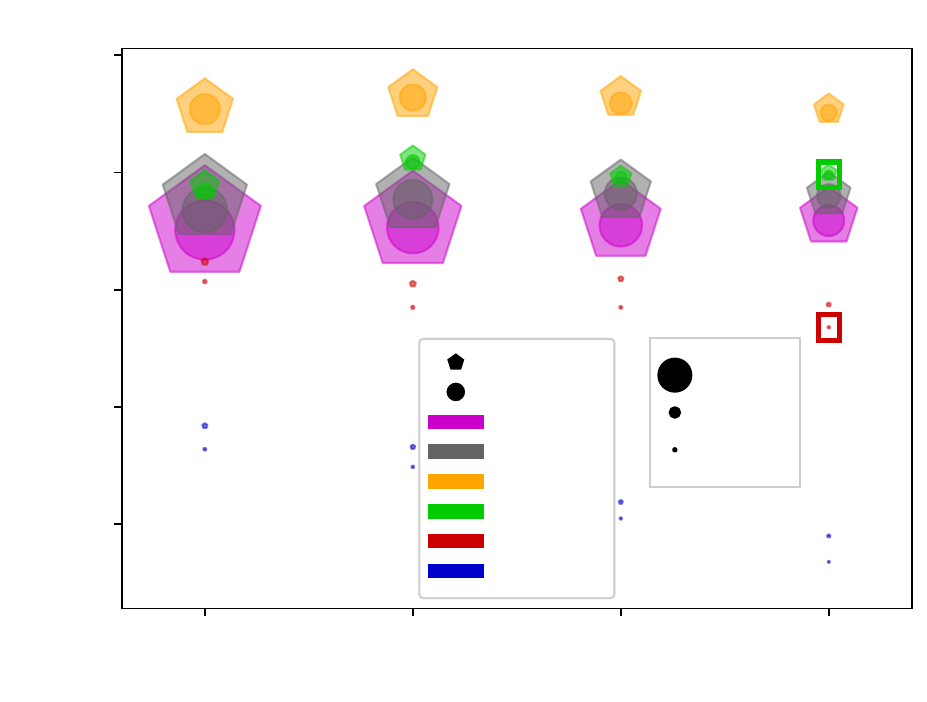}
        \caption{}
        \label{fig:Ablation_d}
    \end{subfigure}
    \caption{Ablation study for different model architectures with different pruning percentages and bit-widths. (a) SNDR for monkey K, (b) R2-score for monkey K, (c) SNDR for monkey L and (d) R2-score for monkey L. Marker size is proportional to the size of the encoder part of the model. The MobileNetV1-CAE(0.25x) and DS-CAE1 models with a bit-width of 8 and 75\% pruning percentage are deployed on RAMAN for FPGA evaluation (highlighted by green and red boxes, respectively).}
    \label{fig:StochAblationStudy}
\end{figure*}

\subsection{Model Performance}
\label{sec:mod_performance}

\begin{table*}
\caption{Comparison between stochastic pruning and standard magnitude-based pruning for 8-bit quantization.}
\label{tab:prune_comp}
\centering
\begin{tblr}{
  row{1} = {c},
  row{2} = {c},
  cell{1}{1-2} = {r=3}{},
  cell{1}{3,9} = {c=6}{c},
  cell{2}{3,6,9} = {c=3}{},
  cell{2}{12} = {c=3}{},
  cell{4,7,10}{1} = {r=3}{},
  vline{1-3,6,9,12,15} = {-}{},
  hline{1-4,7,10,13} = {-}{},
  rows = {rowsep=1pt, valign=m},
  columns = {colsep=4pt},
}
\textbf{Model}              & {\textbf{Sparsity}\\\textbf{(\%)}} & \textbf{Monkey K}           &            &                 &                           &            &                 & \textbf{Monkey L}          &            &                 &                           &            &                 \\
                   &          & \textbf{Stochastic Pruning} &            &                 & \textbf{Magnitude-based Pruning} &            &                 & \textbf{Stochastic Pruning} &            &                 & \textbf{Magnitude-based Pruning}  &            &                 \\
                   &          & {\textbf{\textit{SNDR}}\\\textbf{\textit{(dB)}}}  & {\textbf{\textit{R2}}\\\textbf{\textit{Score}}} & {\textbf{\textit{Size}}\\\textbf{\textit{(kB)$^\dagger$}}} & {\textbf{\textit{SNDR}}\\\textbf{\textit{(dB)}}}  & {\textbf{\textit{R2}}\\\textbf{\textit{Score}}} & {\textbf{\textit{Size}}\\\textbf{\textit{(kB)$^\dagger$}}} & {\textbf{\textit{SNDR}}\\\textbf{\textit{(dB)}}}  & {\textbf{\textit{R2}}\\\textbf{\textit{Score}}} & {\textbf{\textit{Size}}\\\textbf{\textit{(kB)$^\dagger$}}} & {\textbf{\textit{SNDR}}\\\textbf{\textit{(dB)}}}     & {\textbf{\textit{R2}}\\\textbf{\textit{Score}}} & {\textbf{\textit{Size}}\\\textbf{\textit{(kB)$^\dagger$}}}  \\
{\textbf{MobileNetV1-}\\ \textbf{CAE(1.00x)}} & \textbf{25} & {25.23 \\$\pm$ 2.53} & {0.90 \\$\pm$ 0.08} & 2456.384 & {25.20 \\$\pm$ 2.50} & {0.89 \\$\pm$ 0.08} & 3633.728 & {28.22 \\$\pm$ 2.37} & {0.95 \\$\pm$ 0.18} & 2456.384 & {28.19 \\$\pm$ 2.37} & {0.94 \\$\pm$ 0.17} & 3633.728        \\
                   & \textbf{50} & {25.22 \\$\pm$ 2.54} & {0.89 \\$\pm$ 0.08} & 1671.488 & {25.20 \\$\pm$ 2.51} & {0.89 \\$\pm$ 0.08 }& 2456.384 & {28.24 \\$\pm$ 2.38} & {0.95 \\$\pm$ 0.17} & 1671.488 & {28.18 \\$\pm$ 2.40} & {0.94 \\$\pm$ 0.18} & 2456.384        \\
                   & \textbf{75}  & {25.05 \\$\pm$ 2.48} & {0.89 \\$\pm$ 0.09} & 886.592 & {25.14 \\$\pm$ 2.50} & {0.89 \\$\pm$ 0.08} & 1279.04 & {28.27 \\$\pm$ 2.35} & {0.95 \\$\pm$ 0.17} & 886.592 & {28.18 \\$\pm$ 2.41} & {0.94 \\$\pm$ 0.18} & 1279.04         \\
{\textbf{MobileNetV1-}\\ \textbf{CAE(0.25x)}} & \textbf{25} & {24.33 \\$\pm$ 2.21}  & {0.87 \\$\pm$ 0.10} & 173.36  & {23.94 \\$\pm$ 2.02} & {0.86 \\$\pm$ 0.14}  & 246.32  & {28.63 \\$\pm$ 2.34}  & {0.95 \\$\pm$ 0.15}  & 173.36 & {28.48 \\$\pm$ 2.28} & {0.95 \\$\pm$ 0.17} & 246.32 \\
                   & \textbf{50}     & {24.18 \\$\pm$ 2.21} & {0.87 \\$\pm$ 0.13} & 124.72 & {23.91 \\$\pm$ 2.03}  & {0.86 \\$\pm$ 0.14}  & 173.36 & {28.48 \\$\pm$ 2.29} & {0.95 \\$\pm$ 0.14} & 124.72 & {28.55 \\$\pm$ 2.29} & {0.95 \\$\pm$ 0.16} & 173.36  \\
                   & \textbf{75}     & {24.37 \\$\pm$ 2.19} & {0.87 \\$\pm$ 0.10} & 76.08 & {24.03 \\$\pm$ 2.05} & {0.86 \\$\pm$ 0.14} & 100.4 & {28.49 \\$\pm$ 2.28} & {0.95 \\$\pm$ 0.15} & 76.08 & {28.57 \\$\pm$ 2.31} & {0.95 \\$\pm$ 0.15} & 100.4           \\
\textbf{DS-CAE1} & \textbf{25} & {22.78 \\$\pm$ 2.38} & {0.82 \\$\pm$ 0.16} & 10.8 & {22.75 \\$\pm$ 2.35} & {0.81 \\$\pm$ 0.14} & 14.256 & {27.55 \\$\pm$ 2.42} & {0.94 \\$\pm$ 0.15} & 10.8 & {27.78 \\$\pm$ 2.51} & {0.94 \\$\pm$ 0.16} & 14.256          \\
                   & \textbf{50}  & {22.70 \\$\pm$ 2.21} & {0.81 \\$\pm$ 0.19} & 8.496 & {22.69 \\$\pm$ 2.30}  & {0.81 \\$\pm$ 0.14} & 10.8  & {27.55 \\$\pm$ 2.52} & {0.94 \\$\pm$ 0.15} & 8.496 & {27.78 \\$\pm$ 2.53}  & {0.94 \\$\pm$ 0.16} & 10.8            \\
                   & \textbf{75} & {22.61 \\$\pm$ 2.21} & {0.81 \\$\pm$ 0.13} & 6.192 & {22.38 \\$\pm$ 2.18} & {0.80 \\$\pm$ 0.14} & 7.344  & {27.43 \\$\pm$ 2.41}  & {0.94 \\$\pm$ 0.13} & 6.192 & {27.61 \\$\pm$ 2.50} & {0.94 \\$\pm$ 0.15} & 7.344
\end{tblr}

\begin{tablenotes}
  \small
  \item $^{\dagger}$ The parameter size of the encoder part.
\end{tablenotes}
\end{table*}

\begin{figure*}[h]
    \centering
    \begin{subfigure}[]{0.49\textwidth}
        \centering
        \includegraphics[width=\textwidth]{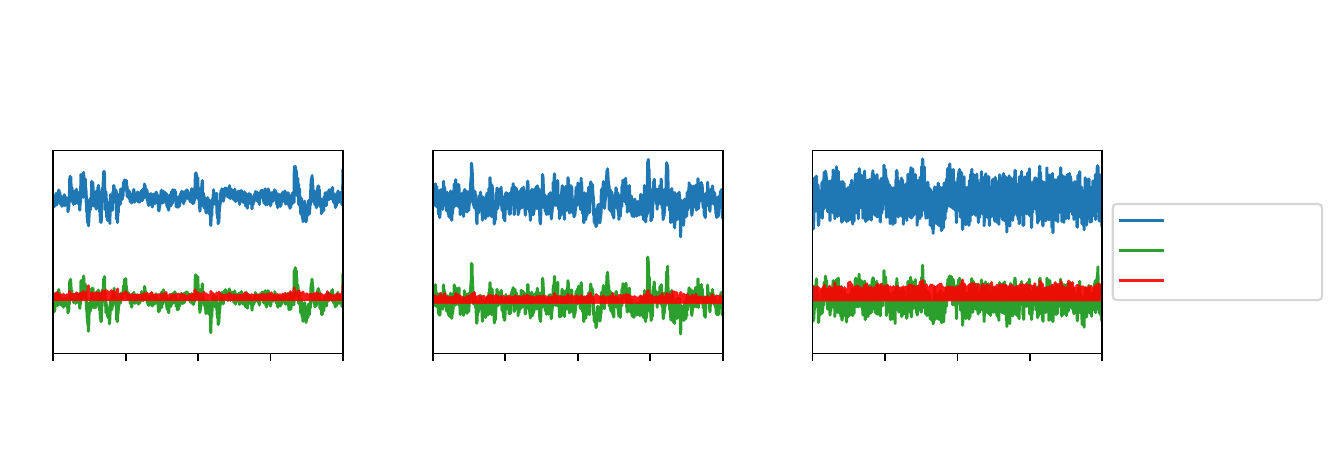}
        \caption{}
        \label{fig:Signals_a}
    \end{subfigure}
    ~
    \begin{subfigure}[]{0.49\textwidth}
        \centering
        \includegraphics[width=\textwidth]{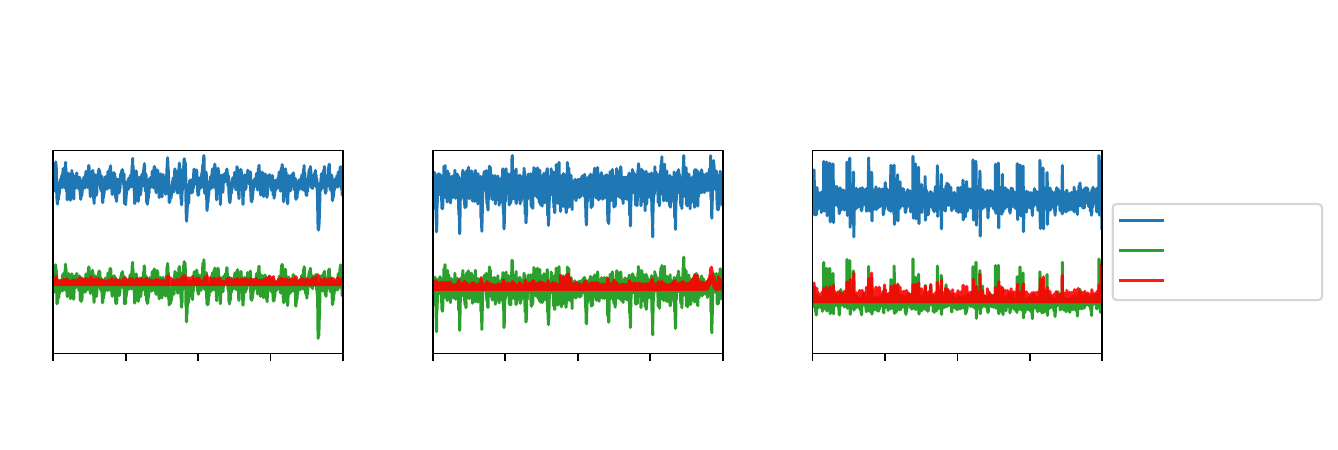}
        \caption{}
        \label{fig:Signals_b}
    \end{subfigure}
    \\
    \begin{subfigure}[]{0.49\textwidth}
        \centering
        \includegraphics[width=\textwidth]{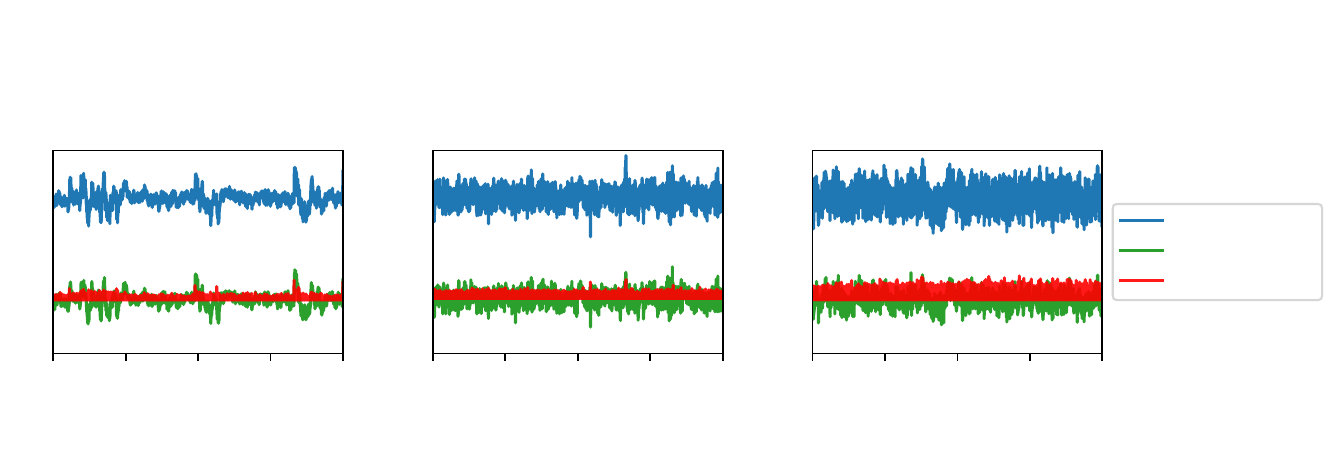}
        \caption{}
        \label{fig:Signals_c}
    \end{subfigure}
    ~
    \begin{subfigure}[]{0.49\textwidth}
        \centering
        \includegraphics[width=\textwidth]{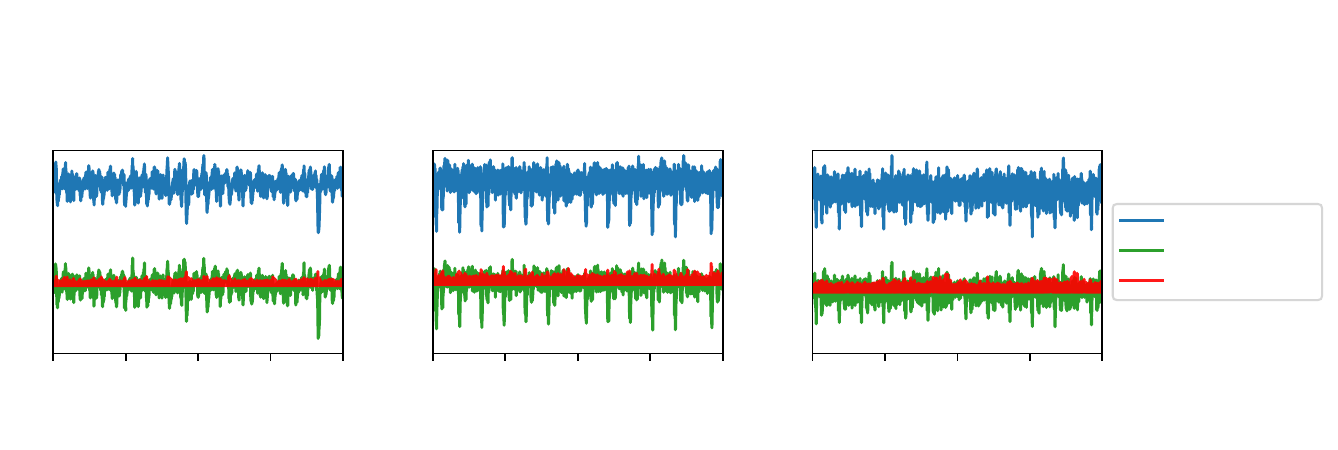}
        \caption{}
        \label{fig:Signals_d}
    \end{subfigure}
    \caption{Original and reconstructed signals along with their absolute differences for MobileNetV1-CAE(0.25x) and DS-CAE1 model using 8-bit quantization and 75\% sparsity. The channels with the best, median, and worst SNDRs are shown.}
    \label{fig:Signals}
\end{figure*}

We experimented with six models. Among them, four are MobileNetV1 \cite{Howard2017} based autoencoder models with width multipliers 1, 0.75, 0.5, and 0.25, and the others are custom depth-wise separable convolutional autoencoder models DS-CAE1 and DS-CAE2. The MobileNetV1-based CAE was chosen as a baseline model for comparison because it inherently employs depthwise separable convolutions, for which RAMAN is tuned and optimized. Additionally, MobileNet significantly reduces the number of parameters and computations in the network, leading to reduced storage, latency, and memory access requirements. This reduction is critical for deployment on edge devices for applications such as BCI. The architectures of the MobileNetV1-CAE model with width multiplier 1 and the two DS-CAE models are presented in Table \ref{tab:mob_1_table} and \ref{tab:ds_cae_table}, respectively. For the MobileNetV1-CAE models with a width multiplier other than 1, the number of channels in each layer is multiplied by the width multiplier and rounded to the nearest integer which is greater than or equal to it and divisible by 16 as shown in \eqref{channel_w_mul}. 
\begin{equation}
N_{l, w} = \left\lceil \frac{N_l \times w}{16}\right\rceil \times 16
\label{channel_w_mul}
\end{equation}  
where $N_{l, w}$ and $N_l$ are the number of channels in layer $l$ for MobileNetV1-CAE model with width multiplier $w$ and 1, respectively.
\par
The PyTorch framework is used to train the models \cite{Pytorch2019}. To demonstrate the benefits of employing stochastic pruning over magnitude-based pruning, we conducted training of the models using two pruning methods separately. The results are depicted in the Table \ref{tab:prune_comp}. For magnitude-based pruning, the baseline floating-point model is trained for 500 epochs. Then, pruning is applied in the order of 25\%, 50\%, and 75\% weight sparsity. After each pruning step, the model is trained for 100 epochs with the corresponding weight sparsity level. In the case of stochastic pruning, since the prune mask is known beforehand for all the sparsity levels, the models are pruned with that specific mask at the beginning and then trained for 500 epochs. The pruned 32-bit floating point models are then quantized to 8-bit weights, and quantization-aware training (QAT) is performed for 50 more epochs using integer-only arithmetic \cite{Jacob2017}. We employed batch normalization folding \cite{Yao2021HAWQV3} to enhance the efficiency of our model, thereby reducing computational overhead while maintaining performance. We trained the models using Adam optimizer and 1cycle learning rate scheduler \cite{OneCycle2018}, setting a maximum learning rate of 0.01 and using a mini-batch size of 128. The mean absolute error (MAE) between the input and reconstructed signals was considered as the loss function.

\begin{table*}
\caption{Comparison of performance of 8-bit quantized models trained on individual$^{\dagger}$ and combined$^{\star}$ dataset.}
\label{tab:combined}
\begin{subtable}[t]{\columnwidth}
\centering
\begin{tblr}{
  row{1} = {c},
  cell{1}{1} = {r=2}{},
  cell{3,5,7,9}{1} = {r=2}{},
  cell{1}{2} = {r=2}{},
  cell{1}{3,5} = {c=2}{},
  vline{1-3,5,7} = {-}{},
  hline{1-3,5,7,9,11} = {-}{},
  rows = {rowsep=1.5pt, valign=m},
  columns = {colsep=5pt},
}
{\textbf{Sparsity}\\\textbf{(\%)}} & {\textbf{Training}\\ \textbf{Dataset}}  & \textbf{Monkey K}  & & \textbf{Monkey L} & \\
&   & {\textbf{SNDR}\\ \textbf{(dB)}} & {\textbf{R2}\\ \textbf{Score}} & {\textbf{SNDR}\\ \textbf{(dB)}} & {\textbf{R2}\\ \textbf{Score}} \\ 
\textbf{0} & Individual & {23.72 \\$\pm$ 2.02} & {0.85 \\$\pm$ 0.14} & {28.40 \\$\pm$ 2.25} & {0.95 \\$\pm$ 0.17} \\
              & Combined & {24.26 \\$\pm$ 2.01} & {0.87 \\$\pm$ 0.13} & {29.46 \\$\pm$ 2.33} & {0.96 \\$\pm$ 0.14} \\
\textbf{25} & Individual & {24.33 \\$\pm$ 2.21} & {0.87 \\$\pm$ 0.10} & {28.63 \\$\pm$ 2.34} & {0.95 \\$\pm$ 0.15} \\
              & Combined & {24.35 \\$\pm$ 2.07} & {0.87 \\$\pm$ 0.11} & {29.52 \\$\pm$ 2.17} & {0.96 \\$\pm$ 0.13} \\
\textbf{50} & Individual & {24.18 \\$\pm$ 2.21} & {0.87 \\$\pm$ 0.13} & {28.48 \\$\pm$ 2.29} & {0.95 \\$\pm$ 0.14} \\
              & Combined & {24.30 \\$\pm$ 2.03} & {0.87 \\$\pm$ 0.15} & {29.55 \\$\pm$ 2.24} & {0.96 \\$\pm$ 0.14} \\
\textbf{75} & Individual & {24.37 \\$\pm$ 2.19} & {0.87 \\$\pm$ 0.10} & {28.49 \\$\pm$ 2.28} & {0.95 \\$\pm$ 0.15} \\
              & Combined & {24.47 \\$\pm$ 1.99} & {0.88 \\$\pm$ 0.13} & {29.49 \\$\pm$ 2.22} & {0.96 \\$\pm$ 0.15}
\end{tblr}
\caption{MobileNetV1-CAE(0.25x)}
\label{tab:mob_25_combined}
\end{subtable}
\begin{subtable}[t]{\columnwidth}
\centering
\begin{tblr}{
  row{1} = {c},
  cell{1}{1} = {r=2}{},
  cell{3,5,7,9}{1} = {r=2}{},
  cell{1}{2} = {r=2}{},
  cell{1}{3,5} = {c=2}{},
  vline{1-3,5,7} = {-}{},
  hline{1-3,5,7,9,11} = {-}{},
  rows = {rowsep=1.5pt, valign=m},
  columns = {colsep=5pt},
}
{\textbf{Sparsity}\\\textbf{(\%)}} & {\textbf{Training}\\ \textbf{Dataset}}  & \textbf{Monkey K}  & & \textbf{Monkey L} & \\
&   & {\textbf{SNDR}\\ \textbf{(dB)}} & {\textbf{R2}\\ \textbf{Score}} & {\textbf{SNDR}\\ \textbf{(dB)}} & {\textbf{R2}\\ \textbf{Score}} \\ 
\textbf{0} & Individual & {22.72 \\$\pm$ 2.35} & {0.81 \\$\pm$ 0.14} & {27.71 \\$\pm$ 2.47} & {0.94 \\$\pm$ 0.17} \\
              & Combined & {22.65 \\$\pm$ 2.24} & {0.81 \\$\pm$ 0.14} & {27.32 \\$\pm$ 2.35} & {0.93 \\$\pm$ 0.17} \\
\textbf{25} & Individual & {22.78 \\$\pm$ 2.38} & {0.82 \\$\pm$ 0.16} & {27.55 \\$\pm$ 2.42} & {0.94 \\$\pm$ 0.15} \\
              & Combined & {22.56 \\$\pm$ 2.17} & {0.81 \\$\pm$ 0.16} & {27.33 \\$\pm$ 2.39} & {0.93 \\$\pm$ 0.17} \\
\textbf{50} & Individual & {22.70 \\$\pm$ 2.21} & {0.81 \\$\pm$ 0.19} & {27.55 \\$\pm$ 2.52} & {0.94 \\$\pm$ 0.15} \\
              & Combined & {22.53 \\$\pm$ 2.15} & {0.81 \\$\pm$ 0.16} & {27.35 \\$\pm$ 2.45} & {0.93 \\$\pm$ 0.15} \\
\textbf{75} & Individual & {22.61 \\$\pm$ 2.21} & {0.81 \\$\pm$ 0.13} & {27.43 \\$\pm$ 2.41} & {0.94 \\$\pm$ 0.13} \\
              & Combined & {22.28 \\$\pm$ 2.03} & {0.79 \\$\pm$ 0.17} & {27.08 \\$\pm$ 2.43} & {0.93 \\$\pm$ 0.20}
\end{tblr}
\caption{DS-CAE1}
\label{tab:dscae1_combined}
\end{subtable}

\begin{tablenotes}
  \small
  \item $^{\dagger}$ Training set of either Monkey K or Monkey L, evaluation is done on the test set of the same monkey.
  \item $^{\star}$ The training set of both Monkey K and L are combined together.
\end{tablenotes}
\end{table*}

\par
The ability of the models to reconstruct the input signals is evaluated using two metrics, viz. signal-to-noise and distortion ratio (SNDR) and R2 score \cite{Valencia2024}. SNDR is defined as:
\begin{equation}
    \text{SNDR}=20\log_{10}\frac{\|x\|_2}{\|x-\hat{x}\|_2}⁡
\end{equation}
where $x$ and $\hat{x}$ are original and reconstructed signals respectively, and $\|\cdot\|_2$ is the L2-norm operator. The R2 score is calculated as:
\begin{equation}
    \text{R2}=1-\frac{\sum_i\left(x_i-\hat{x}_i\right)^2}{\sum_i\left(x_i-\bar{x}\right)^2}
\end{equation}
where $x_i$ and $\hat{x}_i$ are the $i$-th input and reconstructed sample respectively, and $\bar{x}$ is the mean of the input signals.
\par

Fig. \ref{fig:StochAblationStudy} illustrates the ablation results of our neural network models for both 32-bit floating point and 8-bit quantized versions with 0\%, 25\%, 50\%, 75\% weight sparsity. Fig. \ref{fig:Ablation_a} and \ref{fig:Ablation_b} presents the SNDR and R2 scores respectively for monkey K, while Fig. \ref{fig:Ablation_c} and \ref{fig:Ablation_d} present the same for monkey L. We have used different colors to represent different model architectures, and bit-widths are differentiated by varying the marker shapes. The size of the markers is proportional to the respective encoder parameter size. It is observed from the plot that the custom DS-CAE1 model with 8-bit quantization and 75\% sparsity exhibits comparable performance as the MobileNetV1-CAE models despite a 99.95\% reduction in parameter size as compared to the MobileNetV1-CAE(1x) model with 32-bit floating point weights and 0\% sparsity.
\par
Table \ref{tab:prune_comp} presents a comparison between models pruned with stochastic pruning and those pruned with magnitude-based pruning. The analysis shows that both types of pruning result in nearly equivalent SNDR and R2 scores. However, stochastic pruning reduces the parameter size because it does not require storing the compressed weight indices explicitly in memory. For instance, at 75\% sparsity, the memory size reduction for the MobileNetV1-CAE(0.25x) and DS-CAE1 models was 24.2\% and 15.7\%, respectively. The MobileNetV1-CAE(1x) model achieved the highest parameter memory size reduction of 32.4\%.



\par
In Fig. \ref{fig:Signals}, we show the original signals and their reconstructions along with the absolute error using the MobileNetV1-CAE(0.25x) and DS-CAE1 models with 8-bit quantization and 75\% pruning. These reconstructions were obtained for the channels that have the best, median, and worst SNDRs. The input to the models has a dimension of 96$\times$100, and the encoder output of the MobileNetV1-CAE(0.25x) and DS-CAE1 models are 1$\times$1$\times$256 and 1$\times$1$\times$64 (cf. Table \ref{tab:mob_1_table} and \ref{tab:ds_cae_table}), respectively. Therefore, the MobileNetV1-CAE(0.25x) model has a CR of (96$\times$100)$/$256=37.5, while the DS-CAE1 model achieves a CR of (96$\times$100)$/$64=150.

\par
We investigated how CAE models generalize across diverse monkey recordings. Using data from monkeys K and L, we trained CAE models on a combined dataset with 80\% of the recordings from each monkey and evaluated their performance on individual test sets. Table \ref{tab:combined} compares models trained separately on monkeys K and L with those trained on the combined dataset, using MobileNetV1-CAE(0.25x) and DS-CAE1 models across various weight sparsity values. Notably, models trained on the combined dataset showed similar or improved performance compared to those trained on individual datasets, particularly for the MobileNetV1-CAE(0.25x) model.
\par 
Additionally, we tested the performance of models trained on one monkey (e.g., monkey K) and applied to the other (e.g., monkey L). To avoid overfitting, we trained the model with 80\% of the recordings from monkey K (or L) and 5\% from monkey L (or K), then tested it on the remaining data of monkey L (or K). Using the MobileNetV1-CAE(0.25x) model, we observed 13-14\% reductions in SNDR and 17-18\% reductions in R2 score for monkey K, and 8-16\% reductions in SNDR and 3-9\% reductions in R2 score for monkey L across different sparsity levels, compared to the baseline models trained on the combined dataset (80\% of the recording from each monkey). For the DS-CAE1 model, we observed 7-8\% reductions in SNDR and 10-13\% reductions in R2 score for monkey K, and 7-8\% reductions in SNDR and 6-7\% reductions in R2 score for monkey L across different sparsity levels. This suggests that CAEs can learn effectively even when exposed to a small portion of unseen data during training.

\begin{table*}[b!]
\caption{Comparison of the proposed neural signal compression scheme with existing works.}
\label{tab:comparsion_soa}
\centering
 \resizebox{1\textwidth}{!}{ 
\renewcommand{\arraystretch}{1}
{
\begin{tabular}{|c|c|c|c|c|c|c|c|c|c|} 
\hline 
\textbf{} & \begin{tabular}[c]{@{}c@{}}\textbf{Shoaran}\\ \textbf{et al. \cite{Shoaran2014}}\end{tabular} & \begin{tabular}[c]{@{}c@{}}\textbf{Li}\\ \textbf{et al. \cite{NanLi2016}}\end{tabular}  & \begin{tabular}[c]{@{}c@{}}\textbf{Liu}\\ \textbf{et al. \cite{Liu2016}}\end{tabular} & \begin{tabular}[c]{@{}c@{}}\textbf{Park}\\ \textbf{et al. \cite{Park2018}}\end{tabular} & \begin{tabular}[c]{@{}c@{}}\textbf{Khazaei}\\ \textbf{et al. \cite{Khazaei2020}}\end{tabular}    & \begin{tabular}[c]{@{}c@{}}\textbf{Valencia}\\ \textbf{et al. \cite{Valencia2024}}\end{tabular}  & \begin{tabular}[c]{@{}c@{}}\textbf{Turcotte}\\ \textbf{et al. \cite{Turcotte2015}}\end{tabular} & \begin{tabular}[c]{@{}c@{}}\textbf{Shrivastwa}\\ \textbf{et al. \cite{Shrivastwa2018}}\end{tabular}
& \begin{tabular}[c]{@{}c@{}}\textbf{Our}\\ \textbf{Work}\end{tabular} \\ \hline
\textbf{\begin{tabular}[c]{@{}c@{}}Platform\end{tabular}} & \begin{tabular}[c]  {@{}c@{}}ASIC \\ 180-nm  \end{tabular} & \begin{tabular}[c]  {@{}c@{}}ASIC \\ 130-nm  \end{tabular} &  \begin{tabular}[c]  {@{}c@{}}ASIC \\ 180-nm  \end{tabular} &  \begin{tabular}[c]  {@{}c@{}}ASIC \\ 180-nm  \end{tabular} &  \begin{tabular}[c]  {@{}c@{}}ASIC \\ 130-nm  \end{tabular} &  \begin{tabular}[c]  {@{}c@{}}ASIC \\ 180-nm  \end{tabular} & \begin{tabular}[c]  {@{}c@{}}Xilinx \\ Spartan-6 FPGA  \end{tabular} & \begin{tabular}[c]  {@{}c@{}}Xilinx \\ Virtex-7 FPGA  \end{tabular}  & \begin{tabular}[c]  {@{}c@{}}ASIC 65-nm \& \\ Efinix Ti60 FPGA    \end{tabular}  \\ \hline 
\textbf{\begin{tabular}[c]{@{}c@{}}\textcolor{black}{Signal} \\ \textcolor{black}{Type} \end{tabular}} & \textcolor{black}{EEG} & \textcolor{black}{Spike} & \textcolor{black}{LFP} & \textcolor{black}{LFP} & \begin{tabular}[c]  {@{}c@{}}\textcolor{black}{LFP}  \end{tabular} & \textcolor{black}{LFP} & \textcolor{black}{Spike} & \textcolor{black}{ECoG} & \textcolor{black}{LFP} \\ \hline

\textbf{\begin{tabular}[c]{@{}c@{}}Compression \\ Algorithm \end{tabular}}&  CS & CS & CS & \begin{tabular}[c]  {@{}c@{}}DRR + \\ Hufmann Coding  \end{tabular} & DRR & AE & DWT & CS & CAE \\ \hline

\textbf{Precision}& 10b & 10b & 10b & 10b & 10b  & I/P:16b, O/P:10b & 16b & 16b & W: 8b, Act.: 8b  \\ \hline


\textbf{\begin{tabular}[c]{@{}c@{}}Compression \\ Ratio\end{tabular}}  & $\leq$ 16 & 10 & 8-16 & 4.3-5.8 & 2 & 19.2 & 4.17 & $\leq$ 4 & 150$^\dagger$, 37.5$^{\star}$  \\ \hline

\textbf{SNDR (dB)} & 21.8 & N/A & 9.78 & N/A &  N/A & 19 $\pm$ 3 & 17 & N/A &   \begin{tabular}[c]  {@{}c@{}} \begin{tabular}[c]  {@{}c@{}} 22.61\textcolor{black}{$\pm$2.21}$^{\dagger[K]}$,  27.43\textcolor{black}{$\pm$2.41}$^{\dagger[L]}$ \end{tabular} \\ \begin{tabular}[c]  {@{}c@{}} 24.37\textcolor{black}{$\pm$2.19}$^{\star[K]}$,  28.49\textcolor{black}{$\pm$2.28}$^{\star[L]}$ \end{tabular}   \end{tabular}   \\ \hline


\textbf{R2 Score} & N/A & N/A & N/A & N/A &  N/A   &  \begin{tabular}[c]  {@{}c@{}} 0.72\textcolor{black}{$\pm$0.23}$^{[K]}$ \\ 0.93\textcolor{black}{$\pm$0.09}$^{[L]}$ \end{tabular} & N/A &  N/A & \begin{tabular}[c]  {@{}c@{}} 0.81\textcolor{black}{$\pm$0.13}$^{\dagger[K]}$, 0.94\textcolor{black}{$\pm$0.13}$^{\dagger[L]}$ \\ 0.87\textcolor{black}{$\pm$0.10}$^{\star[K]}$, 0.95\textcolor{black}{$\pm$0.15}$^{\star[L]}$ \end{tabular}  \\ \hline
\textbf{\begin{tabular}[c]{@{}c@{}}Area/ch. \\ (mm$^2$)\end{tabular}} & 0.008 & 0.03 &  0.105 & 0.098  &  0.004 & 0.002 & N/A & N/A &   0.0187  \\ \hline

\textbf{\begin{tabular}[c]{@{}c@{}}Power/ch. \\ ($\mu$W)\end{tabular}} & 0.95  & 12.5 & 3.2 & 15.35  &  6.4  & 0.076 & 5k & N/A & 15.1 \\ \hline
\end{tabular}}
}
\begin{tablenotes}
      \small
      \item $^{[K]}$Monkey K recordings, $^{[L]}$Monkey L recordings obtained from dataset \cite{GnodeDataset}.
    \item $^\dagger$DS-CAE1 model, $^\star$MobileNetV1-CAE(0.25x) model
    \end{tablenotes}
\end{table*}

\subsection {Comparison with prior works}
Table \ref{tab:comparsion_soa} compares the proposed neural signal compression scheme employing RAMAN with existing works. \cite{Shoaran2014,NanLi2016,Liu2016,Shrivastwa2018} employ the compressed sensing scheme for data compression. Shoaran et al. \cite{Shoaran2014} propose an analog domain implementation of the CS algorithm supporting a CR of up to 16, achieving an SNDR of 21.8 dB. The design was implemented using ASIC 180-nm process technology with an area of 0.008 mm$^2$ per channel and power of 0.95 $\micro$W per channel. Li et al. \cite{NanLi2016} introduce a Minimum Euclidean or Manhattan Distance Cluster-based (MDC) deterministic compressed sensing matrix for compressing multi-channel neural signals, achieving a CR of 10. The design was fabricated on 130-nm CMOS with a core area of 0.03 mm$^2$ per channel and power 12.5 $\micro$W per channel. Liu et al. \cite{Liu2016} present a fully integrated wireless neural signal acquisition system with an integrated compressed sensing processor fabricated using 180-nm CMOS technology with a power of 3.2 $\mu$W per channel and an area of 0.105 mm$^2$ per channel. They achieve a CR of 8-16 with 9.78 dB SNDR. Park et al. \cite{Park2018} and Khazaei et al.\cite{Khazaei2020} present lossless compression by employing the dynamic range reduction (DRR) technique. Park et al. \cite{Park2018} exploit the spatial and temporal correlation of neural signals to reduce the dynamic range of LFPs. Additionally, Huffman encoding was applied to compress the LFP signals, achieving an overall average CR of 4.3-5.8. The prototype chip was fabricated using 180-nm CMOS technology with an area of 0.098 mm$^2$ per channel and a power of 15.35 $\micro$W per channel. Khazaei et al. \cite{Khazaei2020} propose a lossless data reduction scheme by eliminating spatial redundancy across parallel recording channels, achieving a CR of 2. The design was fabricated using TSMC 130-nm CMOS technology, occupying a silicon area of 0.004 mm$^2$ per channel, and dissipating 6.4 µW of power per channel. Valencia et al. \cite{Valencia2024} propose an autoencoder-based compression digital architecture for the efficient transmission of LFP neural signals. The compression method outlined in \cite{Valencia2024} differs from our proposed approach. They utilize standard autoencoders (AEs) with dense layers to compress the spatial (channels) domain from 96 to 8. In contrast, our method employs a convolutional autoencoder that compresses both spatial and temporal domains, resulting in a higher compression ratio. Additionally, \cite{Valencia2024} utilizes 16-bit input data samples and 10-bit compressed outputs, achieving an overall CR of 19.2 (96*16/(8*10)). Their design was implemented using 180-nm CMOS process technology with an area of 0.002 mm$^2$ per channel. To benchmark our results, we use the same dataset \cite{GnodeDataset} as in \cite{Valencia2024}. Our SNDR and $R^2$ scores demonstrate superior performance, even at a high compression ratio (CR) of 150, with a per-channel area of 0.0187 mm$^2$ and a power consumption of 15.1  $\mu$W per channel for the DS-CAE1 model operating at 2 MHz.

Additionally, several FPGA-based implementations have been proposed in the literature. Turcotte et al. \cite{Turcotte2015} employs a four-level discrete wavelet transform (DWT) to compress neural data. Their system, comprising a spike detection core, threshold estimation core, and wavelet compression, was implemented on a Xilinx Spartan-6 FPGA. This design achieves a CR of 4.17 with an SNDR of 17 dB consuming power of 5 mW per channel. Shrivastwa et al. \cite{Shrivastwa2018} utilize a combination of compressed sensing and neural networks to compress and reconstruct ECoG signals, respectively. Their design, implemented on a Xilinx Virtex-7 FPGA with 285.5k LUTs and 22.18k registers, achieves a CR of up to 4.

In practical BCI applications where subsequent decoding and processing can reasonably tolerate reconstruction and wireless transmission errors, employing lossy methods such as CAE, AE, or CS with a higher CR presents a more feasible approach. Specifically, the proposed CAE-based compression scheme achieves a significantly higher compression ratio than previous methods due to compression in both spatial and temporal domains while maintaining good SNDR and R2 scores. On the contrary, lossless compression schemes involve reducing the dynamic range of neural signals and encoding them using Huffman coding. While this approach reconstructs the original signal accurately, it typically achieves a very low compression ratio. 
\par Achieving a high CR is crucial for next-generation BCI systems to meet the growing data demands of high-density intracortical recordings. In such high-channel-count settings, the resulting data volume can easily exceed the capabilities of communication bandwidth and energy/power budgets of implantable systems. The proposed scheme effectively addresses this challenge by enabling superior data compression without compromising signal reconstruction quality, thereby facilitating the development of scalable, low-power BCI hardware for future clinical and neuroscience applications.
\par
Additionally, various spike compression methods dedicated to high-density brain-implantable microsystems have been proposed \cite{Chen2023,Shaeri2015,Mohan2023,Nekoui2022,Shaeri2020}. Chen et al. \cite{Chen2023} introduced an online neural signal processor (NSP) for spike detection, feature extraction using first and second derivative extrema, and complex spike clustering using the Geo-OSort algorithm. This clustering algorithm involves threshold calculation and Euclidean distance estimation, taking into account the geometric information of the high-density probe to reduce complexity. Their NSP, fabricated using a 22 nm FDSOI CMOS process, achieved a compression ratio (CR) of 982 with an assumption of 10 spikes/s/channel. Shaeri and Sodagar \cite{Shaeri2015} utilized the Discrete Haar Wavelet Transform (DHWT) and spike extraction, proposing a novel data framing scheme to compress the spike waveforms. Their design, fabricated in a 130 nm CMOS process, achieved a CR of 903. Mohan et al. \cite{Mohan2023} employed a neuromorphic approach to spike compression inspired by DVS sensors \cite{Patrick2008}. They used a delta modulator to encode input spike waveforms into ON/OFF pulses, proposing two transmission modes: 'All pulse mode' (APM) and 'Pulse count mode' (PCM), and exploring the trade-offs between them. They also examined address event representation (AER) for asynchronous data transfer to prevent pulse loss due to collisions, achieving a CR of 40.37 at a 62 Hz firing rate per channel with 90\% spike detection accuracy. Nekoui and Sodagar \cite{Nekoui2022} proposed a spike compression scheme through selective downsampling at the implant side, reconstructing the spike offline using third-order polynomial curve fitting. This design, implemented in 130-nm CMOS technology, achieved a CR of 446.5. In another study, Shaeri and Sodagar \cite{Shaeri2020} presented a framework for on-implant spike sorting based on salient feature extraction, maximizing the geometric mean for spike wave-shape isolation. Their systems include an online spike sorting module configured by a shadow spike sorter block on an external module. Generally, spike compression offers superior CR since only the spike events (with an average firing rate of \textasciitilde40-60 Hz) are detected, extracted, compressed, and transmitted. In contrast, our work focuses on compressing local field potentials (LFPs), which are a different signal modality compared to spike waveforms. LFPs are usually low-frequency components of the neural signal (typically <300 Hz), and the information is encoded in the raw signal itself rather than in spikes. Therefore, in our proposed compression scheme, we compress the raw neural signal as opposed to extracting and compressing spike waveforms. The CR achieved by the proposed compression scheme is superior to the existing LFP compression methods \cite{Liu2016,Park2018,Valencia2024,Khazaei2020}.


\section{Conclusions}
\label{sec:conclusions}
This paper introduces a novel neural signal compression scheme employing convolutional autoencoders. The encoder section of the CAE underwent several hardware-software co-optimizations and was subsequently deployed on the RAMAN tinyML accelerator specifically designed for edge computing applications. RAMAN leveraged weight and activation sparsity to reduce latency, memory usage, and power consumption. A novel hardware-aware stochastic pruning technique was employed to address workload imbalance issues across multiple parallel processing elements and reduce the indexing overhead associated with compressed weight storage, resulting in up to a 32.4\% reduction in parameter memory requirements.
\par
Furthermore, since RAMAN inherently supports a wide range of neural network topologies, including standard convolutions, depth-wise convolutions, point-wise convolutions, pooling layers, and dense layers, the encoder of the CAE was constructed based on depth-wise separable convolutional layers to minimize the number of MAC operations.
\par
The proposed CAE-based scheme performs compression in both spatial (channel) and temporal domains, achieving a superior compression ratio of up to 150. The CAE encoder model was pruned using the stochastic pruning scheme and quantized to 8 bits before deployment on the RAMAN tinyML accelerator. RAMAN was implemented on the Efinix Ti60 FPGA with 52.3k XLR cells and 61 DSP units, and the compressed neural data obtained at the output of RAMAN was decoded offline. Compared to recently reported compression algorithms, our scheme achieves superior reconstruction quality, with signal-to-noise and distortion ratios of 22.6 dB and 27.4 dB and R2 scores of 0.81 and 0.94 for the two monkey neural recordings employing a compact custom-designed DS-CAE1 model. Post-layout analysis of the RAMAN encoder in a TSMC 65-nm CMOS process shows a core area of 1.8 mm\(^2\), which corresponds to approximately 0.0187 mm\(^2\) per channel. Operating at a clock frequency of 2 MHz and a supply voltage of 1.2 V, the encoder has an estimated power consumption of 1.45 mW, translating to 15.1 \(\mu\)W per channel for the DS-CAE1 model.

\section{Acknowledgements}
The authors would like to express their sincere gratitude and appreciation to their colleagues, Srikanth Rohit Nudurupati, Chandana D G, Hitesh Pavan Oleti, Anand Chauhan, Shankaranarayanan H, and Ashwin Rajesh for their invaluable help throughout this work.

\bibliographystyle{IEEEtran}
\bibliography{ref.bib}

\end{document}